\documentclass[iop]{emulateapj}

\def\ltsima{$\; \buildrel < \over \sim \;$}
\def\simlt{\lower.5ex\hbox{\ltsima}}
\def\gtsima{$\; \buildrel > \over \sim \;$}
\def\simgt{\lower.5ex\hbox{\gtsima}}
\def\kpc{{\rm\,kpc}}

\def\pc{{\rm\,pc}}

\def\deg{^\circ}

\def\s{\ifmmode \widetilde \else \~\fi}
\def\={\overline}

\def\spose#1{\hbox to 0pt{#1\hss}}

\def\lta{\mathrel{\spose{\lower 3pt\hbox{$\mathchar"218$}}
     \raise 2.0pt\hbox{$\mathchar"13C$}}}
\def\gta{\mathrel{\spose{\lower 3pt\hbox{$\mathchar"218$}}
     \raise 2.0pt\hbox{$\mathchar"13E$}}}
\def\Dt{\spose{\raise 1.5ex\hbox{\hskip3pt$\mathchar"201$}}}    
\def\dt{\spose{\raise 1.0ex\hbox{\hskip2pt$\mathchar"201$}}}    

\def\dotsfill{\leaders\hbox to 1em{\hss.\hss}\hfill}

\def\Gyr{{\rm\,Gyr}}
\def\FeH{{\rm[Fe/H]}}

\newcommand{\ud}{\mathrm{d}}

\shorttitle{A Bayesian search for dwarf galaxies in PAndAS}
\shortauthors{N. F. Martin et al.}

\begin{document}


\title{The PAndAS view of the Andromeda satellite system -- I. A Bayesian search for dwarf galaxies using spatial and color-magnitude information}


\author{Nicolas F. Martin$^{1,2}$, Rodrigo A. Ibata$^1$, Alan W. McConnachie$^3$, A. Dougal Mackey$^4$, Annette M. N. Ferguson$^5$, Michael J. Irwin$^6$, Geraint F. Lewis$^7$, Mark A. Fardal$^8$}
\email{nicolas.martin@astro.unistra.fr}

\altaffiltext{1}{Observatoire astronomique de Strasbourg, Universit\'e de Strasbourg, CNRS, UMR 7550, 11 rue de l'Universit\'e, F-67000 Strasbourg, France}
\altaffiltext{2}{Max-Planck-Institut f\"ur Astronomie, K\"onigstuhl 17, D-69117 Heidelberg, Germany}
\altaffiltext{3}{NRC Herzberg Institute of Astrophysics, 5071 West Saanich Road, Victoria, BC, V9E 2E7, Canada}
\altaffiltext{4}{Research School of Astronomy \& Astrophysics, The Australian National University, Mount Stromlo Observatory, via Cotter Road, Weston, ACT 2611, Australia}
\altaffiltext{5}{Institute for Astronomy, University of Edinburgh, Blackford Hill, Edinburgh EH9 3HJ}
\altaffiltext{6}{Institute of Astronomy, University of Cambridge, Madingley Road, Cambridge CB3 0HA}
\altaffiltext{7}{Institute of Astronomy, School of Physics A28, University of Sydney, NSW 2006, Australia}
\altaffiltext{8}{Department of Astronomy, University of Massachusetts, Amherst, MA 01003, USA}

\begin{abstract}
We present a generic algorithm to search for dwarf galaxies in photometric catalogs and apply it to the Pan-Andromeda Archaeological Survey (PAndAS). The algorithm is developed in a Bayesian framework and, contrary to most dwarf-galaxy-search codes, makes use of both the spatial and color-magnitude information of sources in a probabilistic approach. Accounting for the significant contamination from the Milky Way foreground and from the structured stellar halo of the Andromeda galaxy, we recover all known dwarf galaxies in the PAndAS footprint with high significance, even for the least luminous ones. Some Andromeda globular clusters are also recovered and, in one case, discovered. We publish a list of the 143 most significant detections yielded by the algorithm. The combined properties of the 39 most significant isolated detections show hints that at least some of these trace genuine dwarf galaxies, too faint to be individually detected. Follow-up observations by the community are mandatory to establish which are real members of the Andromeda satellite system. The search technique presented here will be used in an upcoming contribution to determine the PAndAS completeness limits for dwarf galaxies. Although here tuned to the search of dwarf galaxies in the PAndAS data, the algorithm can easily be adapted to the search for any localised overdensity whose properties can be modeled reliably in the parameter space of any catalog.
\end{abstract}


\section{Introduction}

Astronomy has a long tradition of searching for the proverbial needle in the proverbial haystack, made all the more prominent in the current era of large systematic surveys of the night sky. The search for Local Group dwarf galaxies---groupings of stars tracking specific loci in luminosity and color---is a notable such example. As the number of dwarf galaxies has increased in parallel to the discovery of systems with ever fainter apparent luminosities, so have the intricacies of search techniques.

Putting aside the Magellanic Clouds whose proximity and luminosity renders them visible with the naked eye, initial discoveries of Local Group dwarf galaxies were made possible mid-20th century by the systematic observation of the night sky with photographic plates, such as the Palomar Observatory Sky Survey \citep{shapley39,wilson55}, or specific observations targeting the surroundings of the Andromeda galaxy \citep{vandenbergh72}. Although they were carried out for another 50 years, searches in these photographic surveys led to the discovery of a limited number of new dwarf galaxies at the expense of an excruciating search, followed up by epic campaigns of photometric observations to derive the color-magnitude diagram of dwarf-galaxy candidates (e.g. \citealt{karachentsev98}, \citealt{whiting07}). Studies targeting the M31 surroundings were more tractable, yet still limited by the systematics of photographic plates (e.g. \citealt{armandroff98}, \citealt{karachentsev99}).

The advent of large CCD surveys and the homogeneity they entail significantly bolstered our knowledge of Local Group members. Two particular surveys have been key to the significant increase in the number of known dwarf galaxies within a sphere of $\sim1$-Mpc radius: the Sloan Digital Sky Survey (SDSS), mainly for the discovery of Milky Way companions, and the Pan-Andromeda Archaeological Survey (PAndAS) for the discovery of companions to the M31/M33 couple. In addition, the interest of the ongoing Panoramic Telescope and Rapid Response System~1 (Pan-STARRS1) survey for the search for Local Group dwarf galaxies has recently been made evident by the discovery of two new Andromeda dwarf galaxies \citep{martin13a}. Algorithms developed to trawl these surveys for Local Group satellites have been more and more involved with the necessity to dig deeper and deeper into the noise, as well as with the requirement to determine the surveys' completeness limits, a mandatory step to yield meaningful comparisons between observations of dwarf galaxy systems and their model counterparts.

Most of the techniques developed for searches through the SDSS stellar catalog have relied on more or less complex color(-color)-magnitude cuts so as to inspect only the distribution of stars with the broad properties of dwarf-galaxy members \citep{willman02a,belokurov07a,koposov08}, or that of stars which more closely follow fiducial tracks in color-magnitude space \citep{walsh09}. In both cases, the resulting catalog is then either binned, or convolved with a search-kernel, to enable the location of significant stellar overdensities. Care nevertheless needs to be given to misidentified red background galaxies whose clustering can easily create false-positives. The comparison of the algorithmic significance from stellar catalogs with their application to the catalog of galaxies from the same survey, or the application of a detection-area criterion have both been used in the past to limit the number of these false-positives \citep{koposov08, walsh09}.

Despite the improved efficiency of these more involved search algorithms compared to the mere inspection of photographic plates, they are still limited by at least two factors: the use of sharp color-magnitude cuts which effectively give a probability of 0 or 1 for an object to be a potential dwarf-galaxy-member star, and not directly accounting for contaminants through forward-modeling of local catalog objects. In theory, it would seem that the maximum-likelihood-based matched-filter (MF) technique (e.g. \citealt{kepner99}, \citealt{rockosi02}), applied without binning the data, could bypass both these limitations. However, it is too often forgotten that a mandatory requirement for the MF technique is the assumption of a \emph{uniform} contamination; this requirement is seldom validated for dwarf galaxy searches that span large fractions of the MW sky and, therefore, suffer from varying contamination. As a consequence, even though its straight application can lead to some discoveries (e.g. \citealt{grillmair09}), their true significance is difficult to quantify.

In this paper, we embark on a systematic search for dwarf galaxies in the PAndAS survey. So far, all dwarf galaxies discovered in PAndAS have been unveiled as high-significance detections through the use of techniques similar to those described above: as clear overdensities in the distribution of stellar sources compatible with M31 red-giant-branch stars \citep{martin06b,ibata07,mcconnachie08}, or through the application of a MF technique for the faintest ones \citep{martin09,richardson11}. These new dwarf galaxies have already revealed significant clues about the M31 satellite system (e.g. \citealt{ibata13a}). Yet, fainter dwarf galaxies, essential to constrain the faint end of galaxy formation, must remain to be discovered given the low luminosity systems observed around the Milky Way (e.g. \citealt{martin08b}) and hints seen in the M31 outer halo \citep{mackey13}. Moreover, a systematic search will enable us to determine the search completeness limits, a necessary tool to reliably compare observations with predictions from models of galaxy formation in a cosmological context.

The search technique developed here aims at modeling all sources of contamination (background compact galaxies, foreground Milky Way stars, and M31 stellar halo stars) and we place ourselves in a Bayesian framework to forward-model, at once, the distribution of PAndAS stellar sources in \emph{both} spatial and color-magnitude spaces. The structure of the paper is as follows: \S2 is devoted to a quick description of the PAndAS survey and data, and the data preparation for the particular problem at hand; \S3 presents the dwarf galaxy search algorithm and focusses on the ingredients of our family of models, our priors, and the application to the PAndAS data; the findings of the algorithm are detailed and discussed in \S4 before we conclude in \S5. 

Although presented as a dwarf-galaxy-search algorithm, it should be noted that the basics of the algorithm detailed here can be applied to other searches through any catalog with only minimal modifications. As long as one has a family of reliable models in the parameter space of the catalog for the targets of the search, the principles laid out below remain applicable.

\section{Data}
\subsection{The PAndAS survey}
The PAndAS survey is a Large Programme of the Canada-France-Hawaii Telescope conducted with the MegaPrime/MegaCam wide-field camera over the period 2008--2011 (\citealt{mcconnachie09}, McConnachie et al., in prep). It builds on a pilot survey of the Andromeda surroundings observed with the same instrument and set-up from 2003 until 2008 \citep{ibata07,mcconnachie08}. At its completion, the survey encompasses most of the region within a projected radius of $\sim150\kpc$ from M31, and $\sim50\kpc$ from M33. In total, it has a coverage of $\sim390$\,deg$^2$. Each pointing of the 1-deg$^2$ MegaCam camera was observed in two bands ($g$ and $i$) for $3\times15\,$minutes under good to excellent conditions (seeing $\simlt0.8''$ and median seeing values of $0.67''$ and $0.60''$ in the $g$- and $i$-band, respectively).

The details of the data reduction will be presented in an upcoming publication coinciding with the public release of the survey (McConnachie et al., in preparation; see also Ibata et al., in preparation). In short, the data are preprocessed by CFHT through their Elixir system (this includes de-biasing, flat-fielding, and fringe-correcting the data as well as determining the photometric zero points) before they are further processed using a version of the Cambridge Astronomical Survey Unit (CASU) photometry pipeline \citep{irwin01} which was specially tailored to CFHT/MegaCam observations. Here, the astrometry of individual frames is refined and this information is used to register and then stack component images to create the final image products from which the survey catalogs are generated. Finally, the astrometry is further refined, and objects from the catalogs are morphologically classified (stellar, non-stellar, noise-like) before creating the final band-merged $g$ and $i$ products. The catalogs provide additional quality control information, and the classification step also computes the aperture corrections required to place the photometry on an absolute scale. The band-merged catalogs for each field are then combined to form an overall single entry $g$, $i$ catalog for each detected object. In regions of overlapping fields, only the data from the best-seeing field is kept.The photometric calibration is performed from a combination of staggered short MegCam exposures and PS1 photometry \citep{schlafly12} kindly shared by the PS1 Science Consortium.

\subsection{Data preparation}
The work presented in this paper relies entirely on the catalog from objects reliably classified as stars in both bands. All objects are de-reddened using the \citet{schlegel98} $E(B-V)$ maps, scaled with coefficients derived from the SDSS extinction coefficient and the SDSS to MegaCam color equations:

\begin{eqnarray}
g_0 & = & g - 3.793E(B-V)\\
i_0 & = & i - 2.086E(B-V).
\end{eqnarray}

\noindent Here, $g$ and $i$ refer to observed, calibrated magnitudes, and $g_0$, $i_0$ to their de-reddened equivalent which will be used throughout this paper. To avoid unnecessary clutter, the $_0$ subscript will henceforth often be dropped but the reader should remember that we only refer to de-reddened magnitudes.

Despite the systematic coverage of the PAndAS survey, there are holes in the data, stemming from chip gaps, the presence of saturated bright stars, or a few CCD failures. It would be possible to model these holes out in our analysis, but it was deemed easier to fill the holes by duplicating all the information from neighboring regions (see Ibata et al., in preparation, for more details). These artificial areas are nevertheless tagged so they do not lead to spurious detections in our search for dwarf galaxies. Furthermore, regions of high stellar density around M31, M33, NGC~147 \&~185 are removed from the catalog as they correspond to regions in which crowding becomes an issue, or regions for which the hope of finding a new dwarf galaxy is very slim given the overwhelming presence of contamination from these galaxies.

\begin{figure}
\centering
\includegraphics[width=0.8\hsize,angle=270]{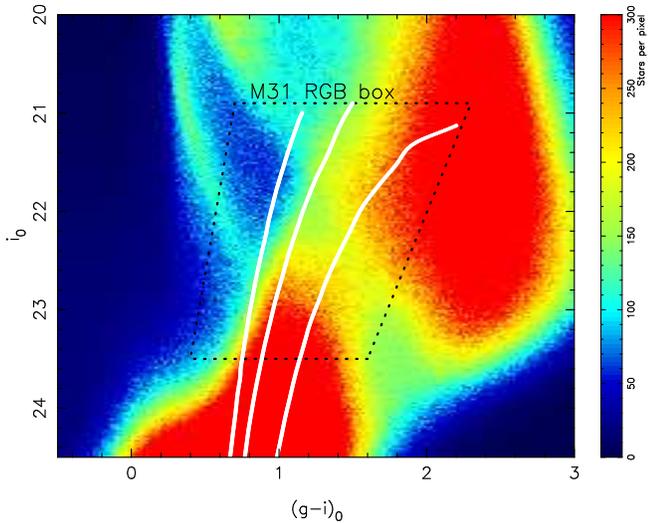}
\caption{\label{total_CMD}The CMD of all stellar sources in the PAndAS survey located away from regions of high density around M31, M33, NGC~147 \&~185. The color coding scales with the density of sources in the $0.02\times0.02$ magnitude pixels. The black dotted line demarcates the M31 RGB box applied to isolate stars of interest for this study. The thick full lines are Padua isochrones \citep{marigo08} at the distance of M31 ($m-M=24.46$), with an age of 13\,Gyr and a metallicity of $\FeH=-2.3$, $-1.4$ and $-0.7$ from left to right.}
\end{figure}


In this contribution, we only care about stars which are potential members of an M31 satellite. A sub-sample of M31 red-giant-branch (RGB) candidate stars is therefore defined to weed out the numerous sources that do not correspond to color and magnitude expectations for old, low-metallicity stars at the distance of M31. This M31 RGB box is shown in Figure~\ref{total_CMD}, overlaid on the color-magnitude diagram of all sources on the PAndAS stellar catalog which are away from M31, M33, NGC~147 \&~185. RGB isochrones of 13\,Gyr-stellar populations with metallicities $\FeH=-2.3$, $-1.4$ and $-0.7$ and a distance modulus\footnote{Throughout his paper, we use the \citet{aconn11,aconn12} favored value for the M31 distance modulus: $m-M = 24.46$. The uncertainties on this measurement are now so small ($\pm0.05$, or $\pm19\kpc$) that taking them into account would have only a negligible impact on our analysis.} $m-M=24.46$ are shown in the same Figure to highlight the expected CMD regions of interest. The M31 RGB selection box is made wide enough so it accounts for increasing photometric uncertainties towards fainter magnitudes, but stops at bright enough magnitudes in both bands to be highly complete for all PAndAS fields. Although a dwarf galaxy located closer to us than M31 would have a fraction of its stars, near the TRGB, that would be brighter than the selection box bright limit, the fraction of these stars is negligible compared to the increasingly numerous stars at fainter magnitudes that will fall in the box. We therefore decide to tailor the selection box to the TRGB at the distance of M31 in order not to add many non-M31 stars to our sample.

Finally, the equatorial coordinates of any star, $(\alpha_k,\delta_k)$, are projected on the plane tangential to the celestial sphere at the location of M31. These projections are referred to as $(X_k,Y_k)$ with X increasing towards the west and Y increasing towards the north.

\begin{figure*}
\centering
\includegraphics[width=0.98\hsize]{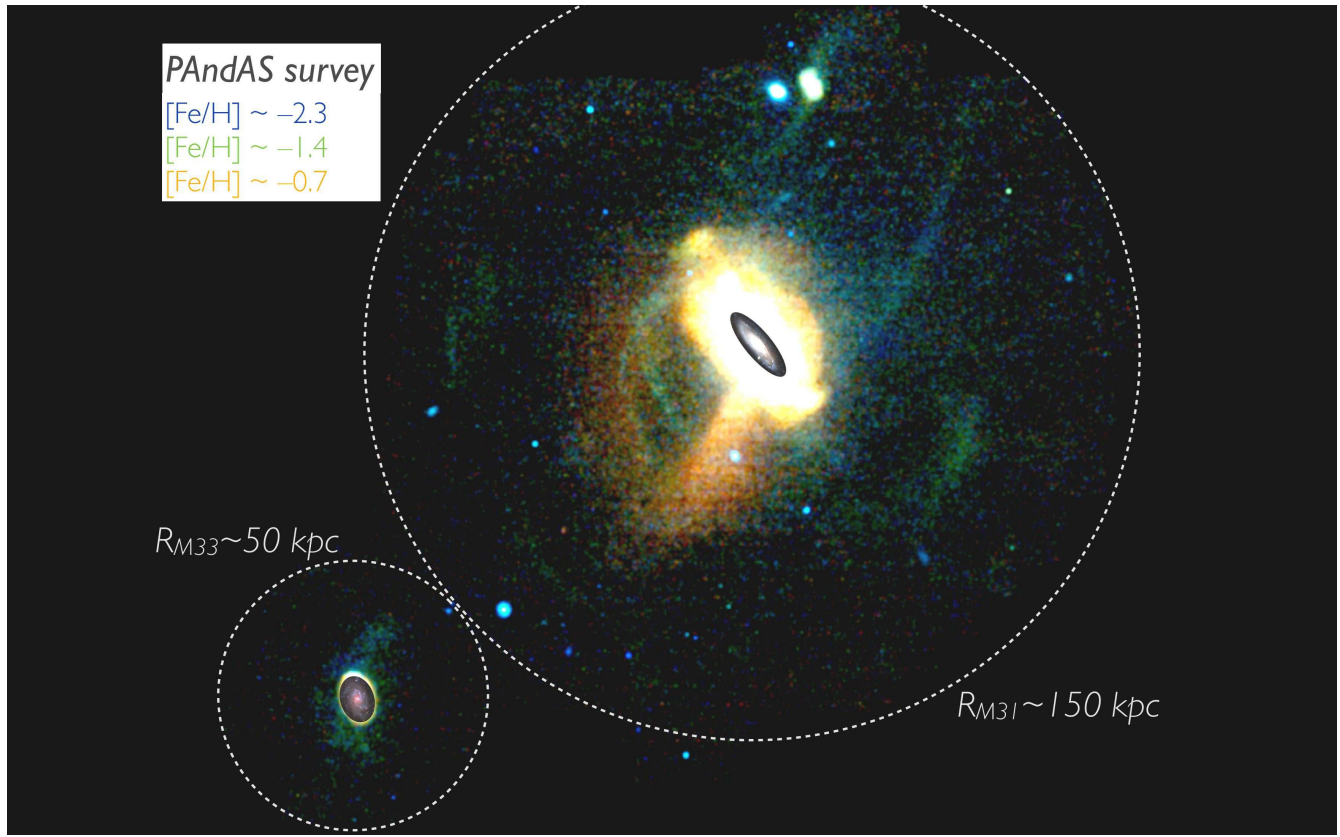}
\caption{\label{M31_PAndAS}Combined Red-Green-Blue color image of the PAndAS survey. Each one of the three channel images is a MF map of the survey for which the signal is the CMD model of an old RGB at the distance of M31, convolved by the observed photometric uncertainties (see \S~\ref{dwarf_galaxy_models}), and the contamination is the spatially varying Milky Way CMD model described in \S~\ref{contamination_models}. The blue, green, and red channels correspond to signal stellar populations of metallicity $\FeH = -2.3$, $-1.4$, and $-0.7$, respectively. All maps have been smoothed with a Gaussian kernel of $2'$ dispersion and the intensity of a pixel scales with the square root of the MF result. The two dotted circles correspond to distances of $\sim150\kpc$ from M31, and $\sim50\kpc$ from M33. The insert images of the M31 and M33 disks illustrate the scale of the survey. The reader is referred to Figure~1 of \citet{lewis13} for the name of the various stellar structures.}
\end{figure*}

In order to provide context, Figure~\ref{M31_PAndAS} presents a mapping of the M31 stellar halo structures. Anticipating slightly on the models that will be presented in section~\S3.2 (and keeping in mind the caveat about the MF technique mentioned in \S~1), this figure shows a combined RGB color image of the PAndAS survey for which the three channel-images are MF maps whose filters are RGB models of metallicity $\FeH=-2.3$ (blue), $-1.4$ (green) and $-0.7$(red), and whose background model corresponds to the contamination model detailed in~\S3.2. The images have been smoothed with a kernel of $2'$ and their luminosity scales as the square root of the matched filter results. The ensuing figure highlights the extent of the M31 stellar halo, as well as the varied nature of the structures, both in shape and in stellar content. Most of the green/blue dots visible on the map correspond to known dwarf galaxies whose discovery was reported in previous contributions \citep{vandenbergh72,armandroff98,zucker04b,martin06b,ibata07,majewski07,zucker07,irwin08,mcconnachie08,martin09,richardson11}.

\section{An automated sky and color-magnitude dwarf galaxy search algorithm}
\subsection{The basics of the method}
\label{section:method}

We wish to search for locations in the PAndAS survey which are likely to be the center of significant overdensities of stars that follow a well-defined density distribution, both on the sky and in color-magnitude (CM) space. For a chosen location, we have a set of $n$ data points, $\mathcal{D}_n = \{\overrightarrow{d_k}\}_{1\leq k\leq n}$, within a spatial neighboring region $\mathcal{A}$. Each datum is defined by its spatial coordinates and a number of magnitude (or color) measurements, $\overrightarrow{d_k} = \{\alpha_k,\delta_k,m_{1,k},m_{2,k},\dots\}=\{X_k,Y_k,(g-i)_k,i_k\}$ for the PAndAS survey. The likelihood that these data points follow a specific overdensity model, defined by the set of parameters $\mathcal{P}=\{p_1,p_2,\dots,p_j\}$, is then defined as

\begin{equation}
\label{eqn:Ptot}
P_\mathrm{tot}\left(\mathcal{D}_n|\mathcal{P}\right) = \prod_k P_k\left(\overrightarrow{d_k}|\mathcal{P}\right),
\end{equation}

\noindent where $P_k\left(\overrightarrow{d_k}|\mathcal{P}\right)$ is the likelihood of datum $k$ to be generated from the model. For the problem at hand, this likelihood can be expressed as the stellar surface density of the model, $\rho_\mathrm{model}$, normalized to the number of stars expected to be in region $\mathcal{A}$. In other words,

\begin{equation}
\label{eqn:Pk}
P_k\left(\overrightarrow{d_k}|\mathcal{P}\right) = \frac{\rho_\mathrm{model}\left(\overrightarrow{d_k}|\mathcal{P}\right)}{\int_\mathcal{A}\rho_\mathrm{model}\ud\mathcal{A}}.
\end{equation}

Following the principles of Bayesian inference, the likelihood of the data given a model, $P_\mathrm{tot}\left(\mathcal{D}_n|\mathcal{P}\right)$, can be related to the important quantity we seek to determine, the probability of a model given the data, $P\left(\mathcal{P}|\mathcal{D}_n\right)$, via our prior knowledge of the model, $P\left(\mathcal{P}\right)$:

\begin{equation}
\label{eqn:bayes}
P\left(\mathcal{P}|\mathcal{D}_n\right) \propto P_\mathrm{tot}\left(\mathcal{D}_n|\mathcal{P}\right) P\left(\mathcal{P}\right).
\end{equation}

The set of parameters which maximize $P\left(\mathcal{P}|\mathcal{D}_n\right)$ define the model favored by the data and, more importantly, the shape of $P\left(\mathcal{P}|\mathcal{D}_n\right)$ around this maximum provides information on the preference of this model compared to others.

\subsection{The family of density models}
The density models we consider are built from the joint contribution of a dwarf galaxy density model, $\rho_\mathrm{dw}\left(\overrightarrow{d_k}|\mathcal{P}^\mathrm{dw}\right)$, and a model of the data contamination, $\rho_\mathrm{cont}\left(\overrightarrow{d_k}|\mathcal{P}^\mathrm{cont}\right)$, such that

\begin{eqnarray}
\label{eqn:rho_model}
\rho_\mathrm{model}\left(\overrightarrow{d_k}|\mathcal{P}\right) = \rho_\mathrm{dw}\left(\overrightarrow{d_k}|\mathcal{P}^\mathrm{dw}\right) + \rho_\mathrm{cont}\left(\overrightarrow{d_k}|\mathcal{P}^\mathrm{cont}\right)\\
\end{eqnarray}

\noindent and $\mathcal{P} = \mathcal{P}^\mathrm{dw} \cup \mathcal{P}^\mathrm{cont}$. We now proceed to describe how the models chosen for $\rho_\mathrm{dw}$ and $\rho_\mathrm{cont}$ were built, and what their parameters, $\mathcal{P}^\mathrm{dw}$ and $\mathcal{P}^\mathrm{cont}$, are.

\subsubsection{Dwarf galaxy models}
\label{dwarf_galaxy_models}
\emph{\textbf{The dwarf galaxy spatial models.}} A family of round exponential sky-projected radial density profiles is chosen to represent the spatial structure of the dwarf galaxies we are searching for. The probability density function (pdf), $P_\mathrm{dw}^\mathrm{sp}$, of a member star to be located at $(X,Y)$, is therefore only dependent on the center of the dwarf galaxy model $(X_0,Y_0)$, as well as its exponential scale-radius, $r_e$, or its half-light radius $r_h=r_e/1.68$:

\begin{eqnarray}
P_\mathrm{dw}^\mathrm{sp}\left(X,Y|X_0,Y_0,r_h\right) = \frac{1.68^2}{2\pi (r_h)^2}\exp\left(-1.68 \frac{r}{r_h}\right),\\
\end{eqnarray}

\noindent with  $r = \sqrt{(X-X_0)^2+(Y-Y_0)^2}$.

With this model, we assume the dwarf galaxies are spherically symmetric. Although this is not the case for faint dwarf galaxies (e.g. \citealt{martin08b,sand12}), the target systems will contain so few stars in PAndAS that any constraint on the ellipticity will be loose at best. Thus, the cost of the added two parameters (the ellipticity and the position angle of the galaxy's major axis) is judged prohibitive for the little added benefit to the final significance of a detection.

\emph{\textbf{The dwarf galaxy CM models.}} In CM space, there is no analytic expression for the pdf of a dwarf galaxy's stars and we therefore rely on the use of theoretical isochrones and luminosity functions. Our initial assumption is that a dwarf galaxy can be approximated by a single stellar population of age $t$, metallicity $\FeH_\mathrm{dw}$, located at a distance modulus $m-M$. For this set of parameters, the probability of having a star at any position of the CM space in the PAndAS bands, $P_\mathrm{dw}^\mathrm{CM}(g-i,i|t,\FeH_\mathrm{dw},m-M)$, is then proportional to the unidimensional line $\mathcal{I}(g'-i',i'|t,\FeH_\mathrm{dw},m-M)$ defined by the isochrone of age $t$, metallicity $\FeH_\mathrm{dw}$, shifted to a distance modulus of $m-M$, and of value the luminosity function of that stellar population at this point, $\phi(i'|t,\FeH_\mathrm{dw},m-M)$, convolved by the survey's two-dimensional Gaussian photometric uncertainties. These have a standard deviation of $\delta_i(i')$ in the $i$-magnitude direction and $\delta_{(g-i)}(g'-i')=\sqrt{\delta_g(g')^2+\delta_i(i')^2}$ in the color direction, yielding

\begin{eqnarray}
\label{eqn:rho_dw_CM}
P_\mathrm{dw}^\mathrm{CM}(g-i,i|t,\FeH_\mathrm{dw},m-M)\hspace{2cm}\\
= \xi \int_\mathcal{I}\phi(i'|t,\FeH_\mathrm{dw},m-M) G\left(i|i',\delta_i(i')\right)\nonumber\\
G\left(g-i|g'-i',\delta_{(g-i)}(g'-i')\right)\nonumber\\
d\mathcal{I}(g'-i',i'|t,\FeH_\mathrm{dw},m-M),\nonumber
\end{eqnarray}

\noindent with $G(x|\mu,\delta) = \frac{1}{\sqrt{2\pi}\delta}\exp\left(-0.5\frac{(x-\mu)^2}{\delta^2}\right)$ and $\xi$ a normalizing constant calculated such that the integral of $P_\mathrm{dw}^\mathrm{CM}(g-i,i|t,\FeH_\mathrm{dw},m-M)$ over the considered region of CM space is unity.

At this point, the dwarf galaxy CM-space models are defined by an age, a metallicity, and a distance modulus. However, the well-known age/metallicity/distance degeneracy at the RGB level makes it unnecessary to keep all three parameters in the analysis as the data are mainly incapable of discriminating between metallicity, age, or distance shifts. Given the consistently old stellar populations present in all the faint dwarf galaxies previously discovered within the PAndAS survey, we consider only a single age of $t=13\Gyr$. In addition, the faint dwarf galaxies so far discovered have severely undersampled RGBs and any constraint on their distance modulus is loose at best. \citet{aconn11,aconn12} determined that the typical distance uncertainty for such a system, although dependent on the amount of contamination from other stellar populations, can reach 100--200\,kpc for the faintest dwarf galaxies. Since this is a sizable fraction of the M31 virial radius and an inadequate model distance modulus will be mainly compensated for by a simple shift in the model metallicity, we fix the distance modulus of the dwarf galaxy models to that of the Andromeda galaxy, $m-M=24.46$. The only parameter that remains for $P_\mathrm{dw}^\mathrm{CM}$ is consequently the metallicity of the dwarf galaxy, $\FeH_\mathrm{dw}$.

As defined above, this model only reproduces the CM pdf of a single stellar population and is therefore likely to be too narrow to represent the thicker, observed RGB of dwarf galaxies. In order to circumvent this difficulty without unnecessarily complicating the models with multiple stellar populations, adding numerous extra parameters that are of little importance to the search of dwarf galaxies but would significantly slow it down, we artificially inflate the photometric uncertainties with an additional photometric scatter term, $\sigma=0.05$, chosen so that the resulting pdf looks similar to the distribution of RGB stars in known PAndAS dwarf galaxies\footnote{For $\FeH\simeq-1.5$, this scatter in color corresponds to a scatter in metallicity of $\sim0.15$~dex, which would of course increase/decrease for lower/higher metallicities.}. In other words, we substitute  $\delta_g(g')$ and $\delta_i(i')$ in the previous equation with

\begin{eqnarray}
\label{eqn:delta_g} &\delta_g(g')= \sqrt{\delta'_g(g')^2+\sigma^2}\\
\label{eqn:delta_i} \textrm{and} & \delta_i(i') = \sqrt{\delta'_i(i')^2+\sigma^2},
\end{eqnarray}
 
 \noindent where $\delta'_g$ and $\delta'_i$ are the observed photometric uncertainties of the data we are modelling. For their expressions, we use the exponential fits to the pre-PAndAS CFHT data photometric uncertainties determined by \citet{ibata07}.\footnote{Although these observations were performed before the formal start of the PAndAS CFHT Large Programme, the observational setup is identical to the one used in PAndAS.} One should in theory further account for the incompleteness of the data in equation~(\ref{eqn:rho_dw_CM}), a field-dependent property, but we circumvent this difficulty by limiting ourselves to the PAndAS magnitude range we know to be homogeneously complete over the survey ($i<23.5$).

For the isochrones and luminosity functions, we use the models of \citet{marigo08}, but note that any reasonable set of isochrones and luminosity functions produce similar significance results for the search, as the algorithm mainly builds on the broad properties of the models. This was tested on a small region with a set of Dartmouth models \citep{dotter08}.

\emph{\textbf{The full dwarf galaxy models.}} Under the final assumption that there are no stellar population radial gradients in the dwarf galaxy models, the sky-projected-color-magnitude joint pdf of a star belonging to the dwarf galaxy model is simply the product of the sky-projected pdf with the CM pdf. For a dwarf galaxy containing $N^*$ member stars, we therefore have

\begin{eqnarray}
\label{eqn:rho_dw}
\rho_\mathrm{dw}\left(X,Y,g-i,i|N^*,X_0,Y_0,r_h,\FeH_\mathrm{dw}\right) \hspace{2cm}\\
= N^* P_\mathrm{dw}^\mathrm{sp}\left(X,Y|X_0,Y_0,r_h\right) P_\mathrm{dw}^\mathrm{CM}(g-i,i|\FeH_\mathrm{dw})\nonumber
\end{eqnarray}

\noindent and $\mathcal{P}^\mathrm{dw}=\{N^*,X_0,Y_0,r_h,\FeH_\mathrm{dw}\}$.

\subsubsection{Contamination models}
\label{contamination_models}
The main difficulty in finding faint dwarf galaxies in the PAndAS footprint comes from the presence of numerous foreground or background contamination sources (foreground faint Milky Way dwarf stars, M31 halo giant stars, unresolved background compact galaxies that appear as point sources), which swamp the minute stellar overdensities that betray the presence of a low luminosity dwarf galaxy. It is therefore essential to have a good contamination model both on the sky and in CM space.

\emph{\textbf{The contamination spatial models.}} In the plane of the sky, we assume that the density of contaminating sources, $\rho_\mathrm{cont}^\mathrm{sp}(X_0,Y_0)$, is locally well approximated by a uniform distribution around any location $(X_0,Y_0)$ in the PAndAS footprint. Although this is clearly not true on large scales given the very structured nature of the M31 stellar halo, this is a good approximation on the scale of dwarf galaxies that subtend sizes of only a few arcminutes.  In order to determine the local density of stars around $(X_0,Y_0)$, the density of sources in the M31 RGB CM-box is calculated for the 36 azimuthally-divided wedges of an annulus centered on $(X_0,Y_0)$, of inner radius $15'$ and outer radius $20'$.  The median of the densities thus measured is taken as the value of $\rho_\mathrm{cont}^\mathrm{sp}(X_0,Y_0)$. Doing so, we are less sensitive to the presence of localized overdensities (e.g. a neighboring dwarf galaxy, globular cluster, or the presence of a nearby stellar stream) than if we were to take the density of the whole annulus.  The inner and outer radii of the annulus have been chosen to be large enough that the annulus is unlikely to contain stars from a faint dwarf galaxy centered on $(X_0,Y_0)$, but to still provide a good representation of the contamination at $(X_0,Y_0)$.

\emph{\textbf{The contamination CM models.}} Determining the pdf of these contaminating sources in CM-space, $P_\mathrm{cont}^\mathrm{CM}$, is a more trying task as it depends strongly on both the type of contaminants, and the location in the survey footprint: foreground Galactic disk stars are much more numerous in the red part of the CM-space than in the blue part, dominated by Milky Way halo stars, whereas M31 stellar halo contaminants are more prominent at faint magnitudes. At the same time, Milky Way (MW) contaminants are more important toward the northern edge of the PAndAS footprint which is closer to the Galactic plane, whereas M31 stellar halo stars are crowding the regions close to M31, and background compact galaxies are present throughout the survey's footprint. We therefore model the pdf of contaminants at any location in CM-space as the sum of a (mainly) MW contamination, $P_\mathrm{cont,MW}^\mathrm{CM}$, and an M31-related contamination, $P_\mathrm{cont,M31halo}^\mathrm{CM}$.

\begin{figure}
\centering
\includegraphics[width=0.9\hsize,angle=270]{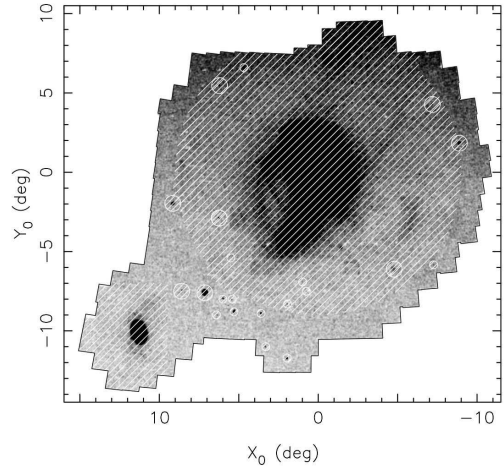}
\caption{\label{contamination_mask}Region $\mathcal{C}$ used to build the MW contamination model. The hashed regions are masked out and not used to fit the contamination model. They correspond to clear stellar overdensities in the PAndAS footprint.}
\end{figure}

\emph{The Milky Way CM contamination models.} The MW contamination pdf is empirically modelled from the data themselves and allowed to vary with the location in the survey. The density map of Figure~\ref{contamination_mask} shows the density of stellar objects in the PAndAS survey which have a color and magnitude compatible with an M31 RGB star.  It reveals the exponential increase of sources to the north (that also happens to mainly correspond to the direction towards the Galactic plane), which is a consequence of the proximity with the MW disk, but this contamination varies differently and is less significant when selecting only very blue stars. The foreground contaminants are expected to either follow an exponential density increase towards the Galactic plane (i.e along the $Y$ axis) for disk stars, as well as a slight increase towards the direction of decreasing $X$ (i.e. mainly decreasing Galactic longitude), or to be reasonably flat over the survey for halo stars and background compact galaxies masquerading as stars. Consequently, we model the density of contaminants at location $(X_0,Y_0)$, at a given color and magnitude $(g-i,i)$, as an exponential increase in both the $X$ and $Y$ directions:

\begin{equation}
\label{eqn:Sigma_contMW}
\Sigma_{(g-i,i)}(X_0,Y_0) = \exp\left(\alpha_{(g-i,i)} X_0+\beta_{(g-i,i)} Y_0+\gamma_{(g-i,i)}\right).
\end{equation}

The functions $\alpha_{(g-i,i)}$, $\beta_{(g-i,i)}$, and $\gamma_{(g-i,i)}$ are determined from a field region, $\mathcal{C}$, of the PAndAS survey which is outside of $9\deg$ (or $\sim120\kpc$ at the distance of M31) from Andromeda's center and further excise any obvious stellar stream or dwarf galaxy, as well as the region around the M33 galaxy (the hashed region in Figure~\ref{contamination_mask}). Although the M31 stellar halo is still present at the edge of the PAndAS survey (Ibata et al. in preparation), such contamination is minimal in these regions. For the center $(g-i,i)$ of each pixel on a fine grid in CM-space, with separations 0.02 in color and magnitude, we proceed to construct a binned spatial map of stellar objects from all stars within a $0.2\times0.2$ magnitude box in CM-space, centered on $(g-i,i)$. A weight map is also built from the map counts by assuming Poisson uncertainties. Using only the map pixels which fall in region $\mathcal{C}$, a weighted $\chi^2$ fit is used to determine parameters $\alpha_{(g-i,i)}$, $\beta_{(g-i,i)}$ and $\gamma_{(g-i,i)}$ for this color and magnitude.\footnote{Allowing for a cross term, $\delta_{(g-i,i)} X_0Y_0$ in the exponential of equation~(\ref{eqn:Sigma_contMW}) almost always results in negligible values of $\delta_{(g-i,i)}$ and does not yield evidently better fits.}
As we repeat this procedure for all the pixels on the fine CM-space grid, we build the functions $\alpha_{(g-i,i)}$, $\beta_{(g-i,i)}$ and $\gamma_{(g-i,i)}$ that track changes with color and magnitude of the contamination over the PAndAS footprint\footnote{The MW CM contamination model, represented by the values of $\alpha(g-i,i)$, $\beta(g-i,i)$, and $\gamma(g-i,i)$ are available by request from the corresponding author.}, as made evident by Figure~\ref{backgd_model_param}. Changes in the strength of these parameters can be linked to astrophysical properties:

\begin{figure}
\centering
\includegraphics[width=0.75\hsize,angle=270]{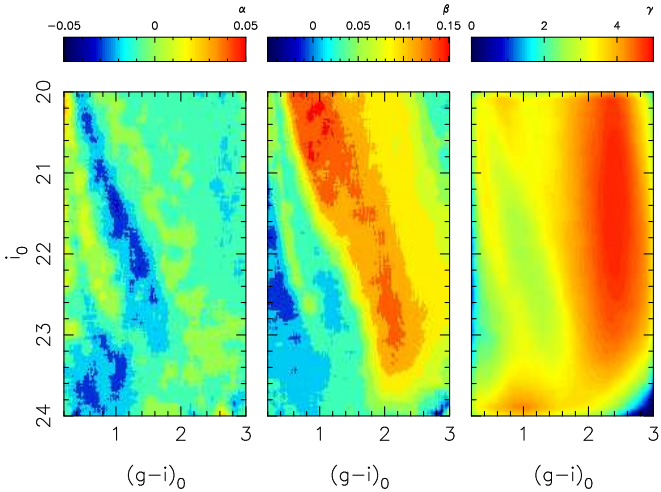}
\caption{\label{backgd_model_param} The MW CM contamination model parameters. From left to right, the panels display the values of parameters $\alpha(g-i,i)$, $\beta(g-i,i)$, and $\gamma(g-i,i)$ from equation (\ref{eqn:Sigma_contMW}), determined as explained in the text.}
\end{figure}

\begin{itemize}
\item $\alpha(g-i,i)$ remains fairly small, as expected from the weak change in the Galactic foreground with Galactic longitude, roughly aligned with the X direction. It nevertheless exhibits an interesting structure of positive $\alpha$ in the regions coincident with MW halo main sequence stars ($0.2<g-i<1.0$ and $i<23.0$), for which $\alpha$ becomes positive, meaning an increase in star counts towards the MW's \emph{anticenter} direction. Although surprising at first, this is confirmed by a detailed study of the complex and structured MW stellar halo over the PAndAS footprint, as will be detailed in another contribution. The impact of these structures on our model are however limited as their stars are mainly bluer than the chosen RGB selection box.
\item $\beta(g-i,i)$ shows more significant changes, with a sharp increase and very positive numbers from $(g-i,i)\sim(0.8,20.0)$ to $(g-i,i)\sim(2.2,23.0)$. This is expected and shows the contamination of foreground MW thick disk main sequence stars whose density increases much more sharply than, for instance, that of MW halo stars. Thin disk stars to the red of this sequence show high but lower values of $\beta$, a likely consequence of these stars being closer, and thus showing a milder density increase towards the Galactic plane.
\item $\gamma(g-i,i)$ corresponds to the normalization of the contamination model and, as such, traces changes in the global density. The corresponding panel of Figure~\ref{backgd_model_param} consequently shows known CMD structures: the MW halo main sequence, the thick disk main sequence, the bulk of foreground thin disk stars piling up in the region $2.0<g-i<3.0$, and even a hint of the population of misidentified background compact galaxies which appears at the faint end of the survey, at $(g-i,i)\sim(1.0,24.0)$.
\end{itemize}

\begin{figure}
\centering
\includegraphics[width=2.1\hsize,angle=270]{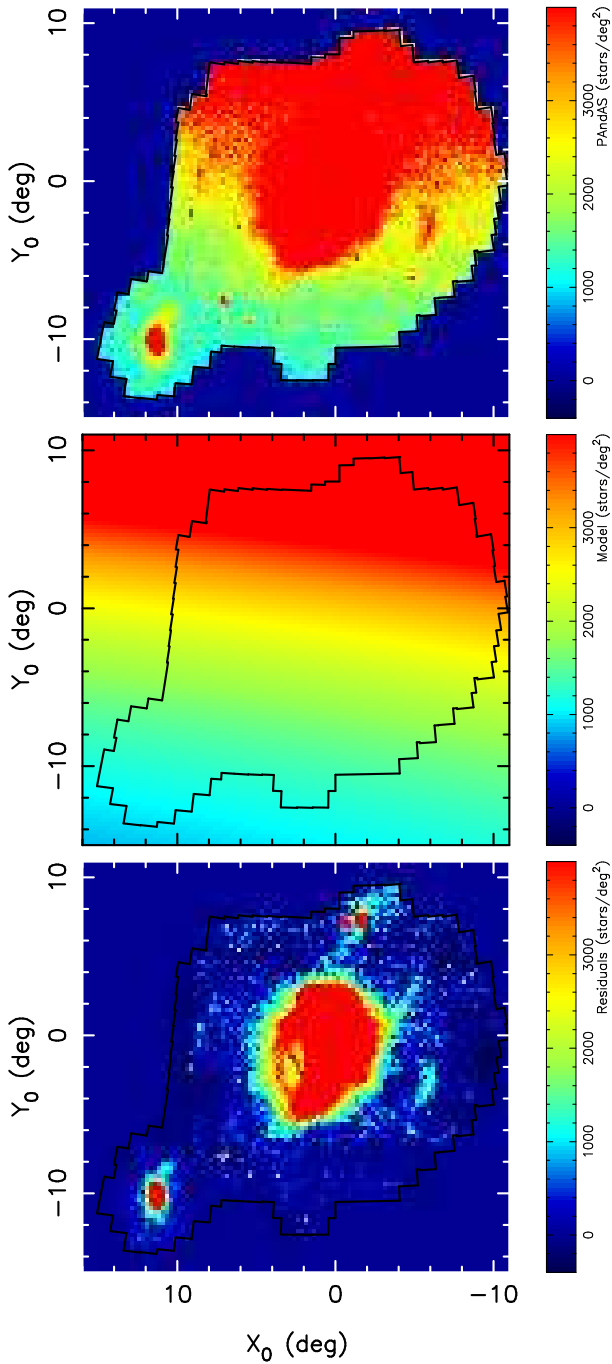}
\caption{\label{contamination_data_model_comparison} \emph{Top panel:} Stellar density of point sources in the M31 RGB box over the PAndAS footprint. M31 is at the center of the map and M33 in visible in the bottom-left corner. Multiple stellar streams also appear in the M31 halo, even though they are dominated by the contamination from the foreground MW. That contamination severely increases towards the north, which mainly corresponds to the axis of increasing Galactic latitude. \emph{Middle panel:} Stellar density of the MW CM contamination model, integrated over the M31 RGB selection box. The model replicates the behavior of the MW foreground contamination visible in the top panel. \emph{Bottom panel:} Residuals between the data and the model, highlighting the fine ability of the model to account for the MW contamination.}
\end{figure}

The power of building such a model for the contamination is that it can now be included in the analytic expression of the global model of stellar populations present at any location of the PAndAS footprint. However, in order to test its quality, it is possible to do a simple and crude subtraction of the contamination model integrated over the M31 RGB box from PAndAS star-count maps. The corresponding maps of the PAndAS data, model, and resulting residuals are displayed in Figure~\ref{contamination_data_model_comparison}. The residual map exhibits a flat and negligible background level over which M31's stellar streams and dwarf galaxies become very prominent, indicative of an adequate modeling of the MW contamination.

\begin{figure}
\centering
\includegraphics[width=0.6\hsize,angle=270]{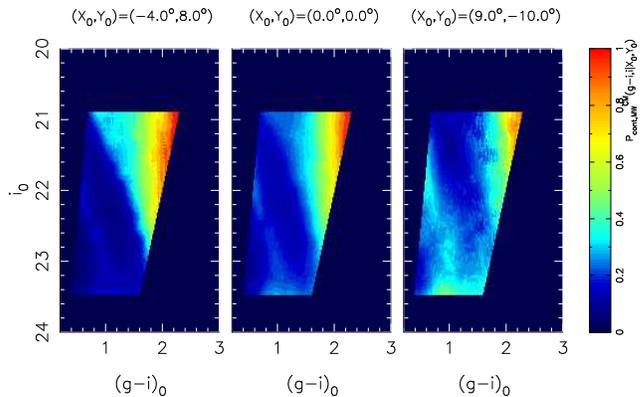}
\caption{\label{sample_Pcont} Realizations of the MW CM contamination pdf, $P_\mathrm{cont,MW}^\mathrm{CM}(g-i,i|X_0,Y_0)$, for 3 arbitrary locations in the survey. The pdfs are normalized to unity over the M31 RGB selection box, leading to changes in the relative importance of the model features.}
\end{figure}

For any location $(X_0,Y_0)$ in the PAndAS footprint, it is now possible to use the values of $\Sigma_{(g-i,i)}(X_0,Y_0)$ at all relevant $(g-i,i)$ and build the contamination model CMD at this particular location. By ensuring that the model is normalized such that the integral of the model is unity over M31 RGB box, it can be interpreted as the pdf of the contamination for this location, $P_\mathrm{cont,MW}^\mathrm{CM}(g-i,i|X_0,Y_0)$. Examples of such pdfs are displayed in Figure~\ref{sample_Pcont} for an arbitrary choice of locations which exemplify the changes in the contribution of respective contaminants (MW thin and thick disks, MW stellar halo,\dots) over the PAndAS footprint. Despite our best effort at only selecting the outskirts of the PAndAS footprint in region $\mathcal{C}$, and carefully masking out stellar streams and dwarf galaxies, there is still a hint of an M31 RGB stellar population in the two right-most panels of Figure~\ref{sample_Pcont}, when the contamination by red, disk stars does not dominate. This feature in the CMDs demonstrates that there is no truly M31-free region in PAndAS (see also Ibata et al., in preparation), but it only has a marginal effect on the search for dwarf galaxies as it effectively means that a small fraction of the M31 stellar halo contamination model (see below) is already accounted for in the MW contamination model; this has little impact on the dwarf galaxy part of the model.

\emph{The M31 CM stellar halo contamination models.} To represent the CM distribution of M31 stellar halo stars, we build isochrone-driven models that are similar to the ones described above for dwarf galaxies, except that we make them broader to account for the multiple stellar populations which contribute to a stellar halo region, likely built from the accretion of multiple dwarf galaxies. The additional photometric scatter term is determined by testing the quality of the search algorithm for increasing value of $\sigma$ in equations~(\ref{eqn:delta_g}) and (\ref{eqn:delta_i}), until most of the M31 stellar streams were not strongly detected for $\sigma=0.15$. $P^\mathrm{CM}_\mathrm{cont,M31}(g-i,i|\FeH_\mathrm{halo})$ is otherwise defined similarly to $P^\mathrm{CM}_\mathrm{dw}(g-i,i|\FeH_\mathrm{dw})$ in equation~(\ref{eqn:rho_dw_CM}). As before, we use the distance/age/metallicity degeneracy to our advantage and only keep a dependence of the isochrones on the metallicity of the halo $\FeH_\mathrm{halo}$ so as not to inflate our model with more parameters that would be poorly constrained. Again, we use an old 13-Gyr-old stellar population and further fix the distance modulus to that of M31 ($m-M = 24.46$).

\emph{The full contamination density model.} As the normalization of the M31 component to the contamination is unknown, we are forced to introduce another parameter, $\eta$, which corresponds to the fraction of the contamination that is in the MW component. Our assumption of a spatially flat contamination density model, whose member stars distribute themselves between a MW contamination model, with fraction $\eta$, and an M31 stellar halo contamination, with fraction $1-\eta$, finally means that

\begin{eqnarray}
\label{eqn:rho_cont}
\rho_\mathrm{cont}\left(g-i,i|X_0,Y_0,\eta,\FeH_\mathrm{halo}\right)\hspace{3.0cm}\\
= \rho_\mathrm{cont}^\mathrm{sp}(X_0,Y_0) \big( \eta P_\mathrm{cont,MW}^\mathrm{CM}(g-i,i|X_0,Y_0) +\hspace{1cm}\nonumber\\
(1-\eta) P^\mathrm{CM}_\mathrm{cont,M31}(g-i,i|\FeH_\mathrm{halo}) \big),\nonumber
\end{eqnarray}

\noindent and that $\mathcal{P}^\mathrm{cont} = \{X_0,Y_0,\eta,\FeH_\mathrm{halo}\}$.

\subsection{Application to the PAndAS data}

The main goal of the search algorithm is to find low luminosity galaxies ($M_V\simgt-8.0$). Even though it will nevertheless find brighter systems with very high significance, we tailor the algorithm to their faint counterparts.

Combining equations~(\ref{eqn:rho_model}), (\ref{eqn:rho_dw}) and (\ref{eqn:rho_cont}), the model density is now entirely expressed as a function of the parameters.  From there, the probability of datum $k$ given these parameters, as expressed in equation~(\ref{eqn:Pk}), will be entirely defined once we have made a choice for the region $\mathcal{A}$ over which we consider neighboring stars to be relevant. This region should be large enough to encompass most of a dwarf galaxy's radial extent, but small enough so the assumption of a constant contamination density is still valid and so little computing time is wasted on determining the probability of stars that are very unlikely to be dwarf galaxy members. $\mathcal{A}$ is therefore chosen to encompass the region within $4r_{h,\mathrm{max}}$ of a dwarf galaxy center $(X_0,Y_0)$, where $r_{h,\mathrm{max}}$ is the largest half-light radius considered when exploring parameter space (see below).

The likelihood of a model is now completely defined and the likelihood of any location of the PAndAS footprint to be at the center of a dwarf galaxy can be evaluated. In order to do so, we start by building a fine grid of centers where the model will be evaluated. This grid needs to be fine enough that we are unlikely to ever be far from the real center of a dwarf galaxy that would be present in the survey. Since M31 dwarf galaxies with $M_V=-6.0$ are expected to have a typical half-light radius of $225\pc$, or $1'$ assuming the distance of M31 \citep{brasseur11b}, we choose a grid space of $0.5'$. This translates into $\sim5$ million centers that are tested for the presence of a dwarf galaxy.

For each center $(X_0,Y_0)$, the search algorithm isolates all the PAndAS stellar sources that fall in the M31 RGB box and within $4r_{h,\mathrm{max}}$ of $(X_0,Y_0)$. These constitute the sample of data points $\mathcal{D}_n$ for this center. Their probability is determined for every 5-tuples of parameters $\mathcal{P}=(N^*,r_h,\FeH_\mathrm{dw},\eta,\FeH_\mathrm{halo})$ selected on a pre-determined 5-dimensional grid. For a choice of parameters, the likelihood of each data point, $P_k\left(\overrightarrow{d_k}|\mathcal{P}\right)$, is calculated by evaluating the dwarf galaxy contribution to the model through equation~(\ref{eqn:rho_dw}), that of the contamination through equation~(\ref{eqn:rho_cont}), then combining them via equation~(\ref{eqn:rho_model}), and finally normalizing the model with equation~(\ref{eqn:Pk}). Multiplying all these probabilities together gives the likelihood of this model, $P_\mathrm{tot}\left(\mathcal{D}_n|\mathcal{P}\right)$, as per equation~(\ref{eqn:Ptot}).

\subsubsection{Priors}
\label{section:priors}
Calculating $P\left(\mathcal{P}|\mathcal{D}_n\right)$ through equation~(\ref{eqn:bayes}) entails making a choice of priors on the 5 parameters which define the model: 

\begin{itemize}
\item We have no clear prior knowledge of the location and density of the stellar streams in the halo of Andromeda. We therefore use a flat prior for $\eta$.
\item We also assume a flat prior on $\FeH_\mathrm{halo}$ and $\FeH_\mathrm{dw}$ over their respective ranges since our assumption of a fixed M31 distance for the CM models would anyway flatten any peaked metallicity prior as distance shifts are compensated for by metallicity shifts.
\item Our prior on the size of the dwarf galaxies is based on the work of \citet{brasseur11b} who determined the size of currently known M31 dwarf galaxies at $M_V = -6.0$ to favor a normal distribution in $\log\left(r_h(\pc)\right)$ with a mean around 2.34 and a dispersion around 0.23. Although there is a slight dependance of the mean of this distribution with the magnitude of dwarf galaxies, this dependance is quite small and we would rather avoid extrapolating this relation to smaller sizes than the range over which it was determined. We therefore assume it to be independent of how populated our galaxy model is over our range of interest ($M_V\simgt-8.0$). To fold this prior into the algorithm, we recalculate the (physical) size distribution to an angular size distribution for the distance modulus of M31 ($m-M = 24.46$).
\begin{figure}
\centering
\includegraphics[width=0.7\hsize,angle=270]{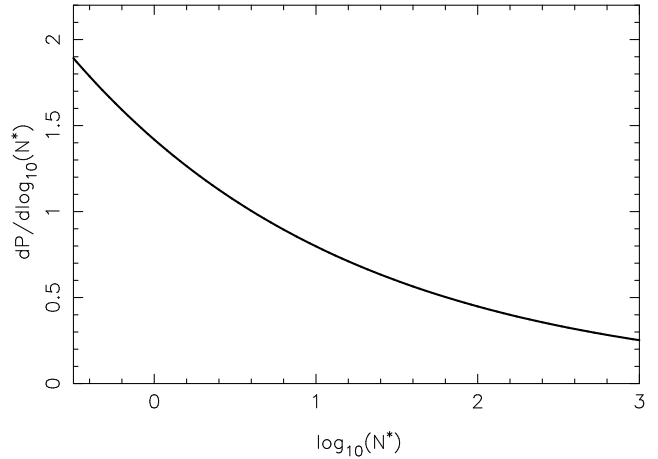}
\caption{\label{prior_logN} The pdf of the prior on $\log_{10}(N^*)$, calculated from the luminosity function of mainly MW satellites determined by \citet{koposov08}.}
\end{figure}
\item The number of stars in the dwarf galaxy part of the model that fall within the M31 RGB selection box, $N^*$, is a quantity that is indirectly related to the magnitude of the dwarf galaxy, via the magnitude limits of the observations. Therefore, the prior on $N^*$ or, more practically, $\log_{10}(N^*)$, is a re-expression of the dwarf galaxy luminosity function. We assume the luminosity function of \citet{koposov08}, determined from the averaged properties of MW and M31 satellites at the bright end, and the MW satellites at the faint end ($dP/dM_V \propto 10^{0.1(M_V+5)}$),\footnote{We further substitute $M_i$ for $M_V$ in this expression under the assumption that faint dwarf galaxies all have similar stellar populations---old and metal-poor---and therefore similar $V-i$ colors.} and convert it to a pdf over $\log_{10}(N^*)$. Although the \citet{koposov08} luminosity function likely differs in its details from the true M31 dwarf galaxy luminosity function, it should nevertheless be a good enough approximation for the problem at hand. The resulting prior is presented in Figure~\ref{prior_logN} and works in bringing down the probability of populated detections by a factor of a few compared to sparsely populated dwarf galaxies.

If a flat prior is used instead, the algorithm yields higher likelihood values for many detections with large favored radii. These are due to the algorithm attempting to model local, slightly non-flat distributions of M31 stellar halo stars by a significant dwarf galaxy with a large $N^*$. It therefore yields an unrealistically large population of populated (i.e. bright) dwarf galaxy detections that are incompatible with our knowledge of dwarf galaxy luminosity functions. The prior on $N^*$ curtails this issue.
\end{itemize}

\subsubsection{Parameter grid}
Although we would prefer using an approach that does not rely on a grid over parameter-space on which to evaluate the models, the sheer number of locations for which parameter space needs to be explored imposes the use of a coarse grid so the calculations are tractable and one can still get a feel for the significance of changes in $P\left(\mathcal{P}|\mathcal{D}_n\right)$ over parameter space. Consequently, we set up a fixed grid over the 5-dimensional space, tested and tailored to focus on the parameter ranges which are the most relevant.

In detail:
\begin{itemize}
\item The fractional contribution of the MW foreground to the total contamination, $\eta$, is sampled from 0.0 to 1.0, with steps of 0.1.
\item $\FeH_\mathrm{halo}$ samples an irregular grid, chosen to span most of the range of stream metallicities in the halo of M31 and therefore emphasizing intermediate metallicities at values of $\FeH=-1.7$, $-1.3$, $-1.1$, $-1.0$, $-0.9$, $-0.8$, $-0.7$, and $-0.6$\,dex. The smaller $\FeH$ steps at higher metallicities are chosen to compensate the increasingly larger $g-i$ color-changes. Although some structures of the M31 halo are more metal-rich than our high limit (e.g. the core of the Southern Giant Stream; \citealt{ibata07}), most of the very metal-rich stars are in fact too red for the M31 RGB box tailored for the metal-poor dwarf galaxies.
\item $\FeH_\mathrm{dw}$ samples a regular grid at the metal-poor end of the metallicity range, from $\FeH=-2.3$ until $\FeH = -1.1$, with steps of 0.3\,dex.
\item $r_h$ is tested from $0.5'$ to $4.0'$ with steps of $0.7'$, spanning the range of most Andromeda satellites, except for the larger ones. Going beyond $r_h=4.0'$ would also be a waste of time given how penalizing our prior on $r_h$ is for such systems. Although there are known M31 dwarf galaxies with half-light radii larger than $4.0'$ \citep{mcconnachie12,martin13a}, the algorithm is tailored towards finding faint galaxies ($M_V\simgt-8.0$), which are likely to be smaller than this limit \citep{brasseur11b}. The presence of a large dwarf galaxy will likely still be detected for the sampled range of $r_h$, as demonstrated by the high significance of the And~XIX detection (see below).
\item $N^*$ is sampled regularly on a $\log_{10}$ scale from $\log_{10}(N^*)=-0.5$, until $\log_{10}(N^*)=3.0$, with steps of 0.5\,dex. The reason we test for models of dwarf galaxies containing less than 1~star in the M31 RGB selection box is related to our evaluation of the significance of a detection and will become apparent in the next sub-section. The high limit of $N^* = 1000$ stars is chosen as it corresponds to a dwarf galaxy with $M_i \simeq -9.5$, which generally would have already been discovered by earlier methods in the PAndAS footprint.
\end{itemize}

\subsubsection{The significance of a detection}
Once the probability of every model parameter 5-tuple has been calculated for a given dwarf galaxy center $(X_0,Y_0)$, it is trivial to find the set of parameters $\widehat{\mathcal{P}}$ that determines the model favored by the data as the set of parameters yielding the highest value of $P(\mathcal{P}|\mathcal{D}_n)$. However, more important is the knowledge of whether this model is significantly better than the favored model containing no dwarf galaxy. This information hinges on the probability of the model, marginalized over all parameters but $\log_{10}(N^*)$,

\begin{eqnarray}
P_{N^*}\left(\log_{10}(N^*)\right)\hspace{5.5cm}\\
= \int_\mathrm{grid} P(\mathcal{P}|\mathcal{D}_n)\ud\eta\ud\FeH_\mathrm{halo}\ud\FeH_\mathrm{dw}\ud r_h.\nonumber
\end{eqnarray}

The significance of a detection is assigned as it would be if $P_{N^*}$ were a Gaussian function. In this case, the favored model of probability $\max(P_{N^*})$ deviates from the model with no dwarf galaxy by $S$ times its dispersion (i.e. an ``$S$--sigma detection'') for $S$ defined as

\begin{equation}
\label{eqn:significance}
S^2 =2\ln\left(\frac{\max(P_{N^*})}{P_{N^*}(N^*=0)}\right).
\end{equation}

In our setup, we explore a grid in $\log_{10}(N^*)$, and we therefore cannot technically calculate $P_{N^*}(N^*=0)$ in equation~(\ref{eqn:significance}). It is replaced by the probability at the lower bound of the grid, $\log_{10}(N^*)=-0.5$, a small enough number for $N^*$ that it has a minimal impact on the calculation of the significance.

\section{Results of the automated search}
\subsection{A detailed view of the And XI-XIII region}
\begin{figure*}
\centering
\includegraphics[width=0.65\hsize,angle=270]{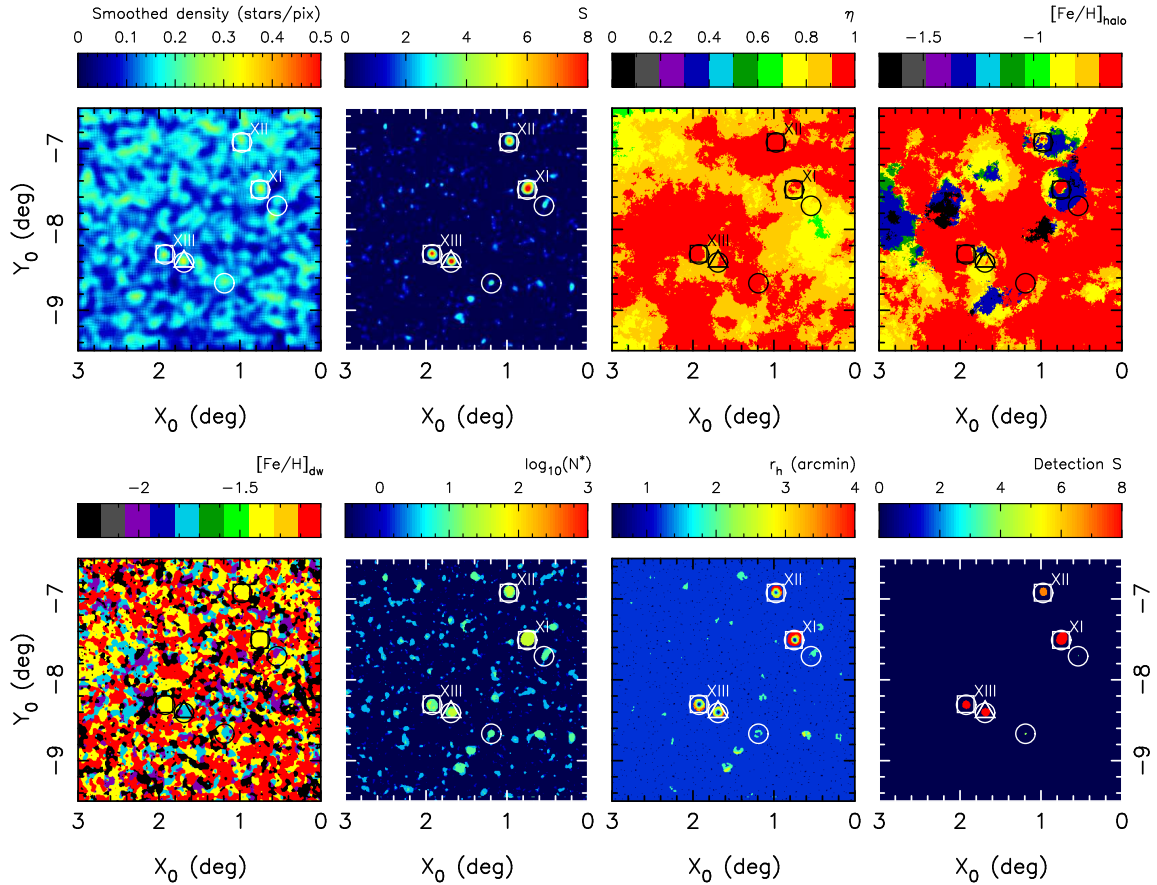}
\caption{\label{results_AndXI-XIII} Results of the search algorithm for the region around And~XI-XIII. From top to bottom, and from left to right, the panels represent the smoothed distribution of point sources in PAndAS; the map of significance $S$; the map of most probable contamination fraction, $\eta$; the map of most probable halo metallicity, $\FeH_\mathrm{halo}$; the map of most probable dwarf galaxy metallicity, $\FeH_\mathrm{dw}$; the map of most probable number of stars in the dwarf galaxy part of the model, $\log_{10}(N^*)$; the map of most probable half-light radius, $r_h$; and the map of detections above the threshold, $S_\mathrm{th}=3.5$, chosen for this region. In all panels, the white or black squares represent the location of the known dwarf galaxies And~XI, XII and~XIII; the white or black triangle represents the location of MGC1. The detections above the threshold are highlighted by white or black circles and include all three dwarf galaxies and the globular cluster.}
\end{figure*}

Before moving to the results of the algorithmic search applied over the whole PAndAS survey, let us first focus on the region around three known dwarf galaxies, And~XI, XII, and~XIII, and study the detailed output of the algorithm for and around these faint systems. The first two panels in the top row of Figure~\ref{results_AndXI-XIII} compare the original smoothed stellar density with the significance of detections for every pixel in the region considered. It is readily evident that the significance map is a huge improvement over a blind search for overdensities in the smoothed density map of all RGB stars. The three known dwarf galaxies in the region, as well as a bright globular cluster (MGC1; \citealt{martin06b}), produce four very significant detections. Even And~XII, one of the faintest known M31 satellites with a total magnitude of $M_V\simeq-6.5$, corresponds to a detection with $S=6.9$. Moreover, supposedly empty M31 halo regions around the four prominent detections invariably have low  significance with at most a few isolated detections with $S=3.0-3.9$. This is a very promising output, emphasizing the efficiency of the algorithm as well as the usefulness of using both the spatial and CMD information when searching for dwarf galaxies.

One could wonder why MGC1 is also found with high significance by the algorithm as it is a compact globular cluster. Despite the core of the cluster being unresolved, the high luminosity of this specific cluster, as well as its very extended halo \citep{mackey10}, lead to tens of outer cluster stars being resolved in the PAndAS survey and forming a reasonably large overdensity of RGB stars, not unlike the dwarf galaxies the algorithm is tailored to find. This is a rather exceptional case as the MegaCam pixel size ($0.187''$) and the seeing are usually too large to allow for resolved M31 globular clusters but we will see in the next section that there is a handful of globular and extended clusters (GCs and ECs) that are found with compelling significances.

The other maps of Figure~\ref{results_AndXI-XIII} present the parameters of the most probable model for each pixel. The fraction of the MW contribution to the overall contamination model, $\eta$, remains consistently high, as expected for an outer halo region with a low density of M31 RGB stars (see Figure~\ref{M31_PAndAS} and Ibata et al., in preparation). As a consequence, most of the panel showing the favored halo metallicity, $\FeH_\mathrm{halo}$, is not well constrained, or simply noise. However, it shows some interesting variations where the fraction of the M31 contamination is non-zero. In particular, one can see variations in the M31 stellar halo metallicity\footnote{One has to remember here that a fixed distance modulus and age was assumed and that these variations could also reflect changes in the distance to the stellar populations, and/or their age.} for potential stellar structures, likely hosting different stellar populations (Ibata et al., in preparation). The lack of significance of most of the region's pixels means that the favored dwarf metallicity, $\FeH_\mathrm{dw}$, varies a lot over the region since it is meaningless for models with low $S$. There are, however, noticeable regions of contiguous pixels at the location of the three dwarf galaxies and MGC1 for which $\FeH_\mathrm{dw}$ becomes well defined at the preferred value for this dwarf galaxy (and the assumption of $m-M=24.46$).

The last two panels show the maps for the two parameters which are directly impacted by the priors we chose. The number of stars in the dwarf galaxy part of the model is dominated by small numbers, as expected from the prior if there is no significant overdensity of stars that would follow our dwarf galaxy model. One can also notice that the few detections with the largest significance also have a non-negligible number of stars in their preferred model. The algorithm favors models with $\log_{10}(N^*) \simeq 1.5$, or about 30 stars, although one should remember that $\log_{10}(N^*)$ is sampled with steps of 0.5, meaning that the numbers of stars within the RGB box in these dwarf galaxies is likely to be in the 20--65 range, in agreement with observations \citep{martin06b}.

Finally, the map for the favored $r_h$ behaves similarly to that for $\log_{10}(N^*)$ in that regions with no significant detection have a preferred size driven by the peak of our prior, around $1'$. Close to significant detections, the size of the preferred model tends to increase at the same time as the significance rises. This is an edge effect as the algorithm starts to feel that there is a significant number of M31 RGB-like stars close to the considered pixel but, since this pixel is not at the center of the dwarf galaxy, it finds them at odds with the assumed, peaked radial distribution. The situation is mitigated by favoring an extended model, as flat as possible. One can also notice a similar effect in the map for $\eta$ and $\FeH_\mathrm{halo}$, for which the algorithm mitigates for the close presence of an overdensity of likely M31 RGB stars by including them in the M31 halo contamination. Overall, the impact of these effects is negligible in our study since all pixels near a significant detection are bundled together and considered as parts of the same detection.

\begin{figure}
\centering
\includegraphics[width=0.8\hsize,angle=270]{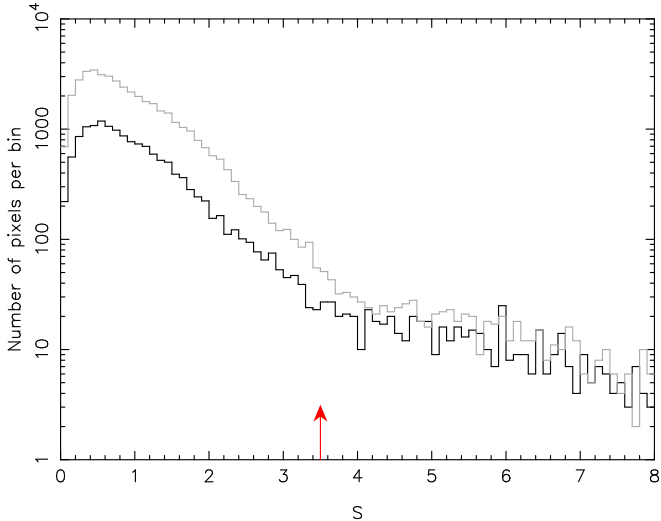}
\caption{\label{significance_hist_AndXI-XIII} Significance distribution for the region around And~XI-XIII. The black line represents the histogram of pixel significance values for the final algorithm whereas the gray histogram corresponds to the significance values when the priors on $\log_{10}(N^*)$ and $r_h$ are substituted with a flat prior. In the latter case, the tail of high significance values becomes evident at larger $S$ values. The threshold signal value, $S_\mathrm{th}$, chosen for this region is shown by the red arrow.}
\end{figure}

If the model were a perfect representation of the data, the distribution of pixel significances would be Gaussian-like for non-negligible significance values, and a tail at high-significance values should be produced by the presence of dwarf-galaxy-like systems in the data. It is this tail of high-significance pixels we wish to identify. The distribution of pixel significances for the region around And XI-XIII is displayed in Figure~\ref{significance_hist_AndXI-XIII} as the black histogram and, except for low significance values for which our significance criterion breaks down, the distribution indeed has a Gaussian behavior until it reaches an almost flat level at the high-significance end. The transition between the two regimes is well marked and guides us in the choice of a detection threshold over which pixels are considered to be worthwhile. Although the exact location of the threshold is not very important, a threshold that is too low will secure more potential discoveries, but also yield many detections that are mere noise fluctuations in the data. Conversely, a threshold that is too high will only yield secured dwarf galaxy detections but will be of little use in our search for fainter dwarf galaxies than the ones we already know. For this region, the high-significance tail starts to deviate from the exponential decline of the distribution for values larger than $S\sim3.5$, a value we use as the detection threshold, $S_\mathrm{th}$, for this region (red arrow).

Figure~\ref{significance_hist_AndXI-XIII} also shows the impact of the priors on the significance, with the gray histogram being produced if we substitute the priors on $\log_{10}(N^*)$ and $r_h$ with flat priors. In an effort to compensate for the not-quite locally flat M31 stellar halo contamination, the algorithm then finds many mildly significant detections that are as flat as allowed by the considered dwarf galaxy models, i.e. a preferred $r_h$ as large as possible and a non-zero preferred $N^*$. These artificial detections (which are not favored when the priors of sub-section~\ref{section:priors} are applied) stretch the significance distribution to larger values, and consequently swarm parts of the real detections. 

Once the detection threshold is selected, the production of a list of detections is only one step away, namely after dealing with the fact that, as can be seen in the second, top-most panel of Figure~\ref{results_AndXI-XIII}, more than a single pixel has a high significance value near a detection. In order to bundle together pixels belonging to the same detection, we order the pixels by significance and, starting with the most significant pixel, we recursively include in a detection all the pixels that are within a radius of 20 pixels (or $10'$) of this current most significant pixel. All the pixels added to a detection are not considered further, whereupon the algorithm moves to the next most significant pixel above the detection threshold and reiterates the process.

\begin{figure}
\centering
\includegraphics[width=0.75\hsize,angle=270]{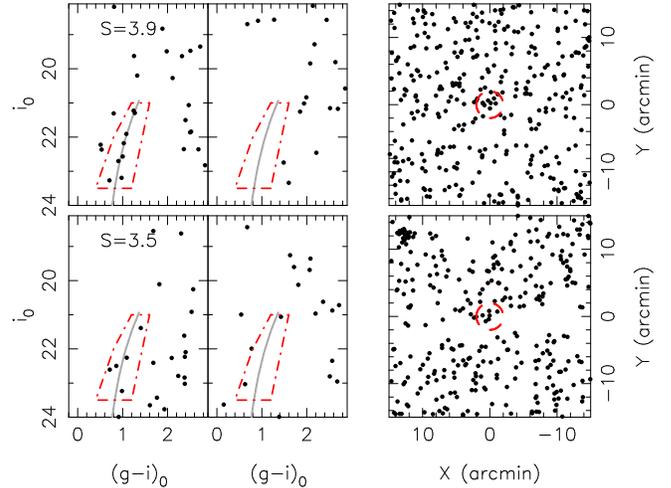}
\caption{\label{AndXI-XIII_detection_CMDs} \emph{Top panels:} The CMD of a region within $2'$ of the most significant unknown detection ($S=3.9$) in the region of And~XI-XIII, along with that of a field annulus of identical coverage at $15'$. The thick, gray line corresponds to a Padua isochrone of age 13 Gyr and $\FeH=-1.7$ at the distance of M31. A group of stars could likely correspond to the RGB of a faint stellar population. The map of stars that fall in the dot-dashed red polygon, centered on the detection, is shown on the right-hand panel. Again, a small grouping of stars is visible. \emph{Bottom panels:} Similar plots for the least significant of the two detections ($S=3.5$). The presence of a compact stellar overdensity is here less obvious, as expected from the lower signal. The grouping of stars to the north east of this detection is And~XI.}
\end{figure}

The resulting detections for the region around And~XI-XIII are shown in the bottom-right-most panel of Figure~\ref{results_AndXI-XIII}, color-coded by the significance of their most significant pixel. Six detections are automatically found: the four known and obvious ones produced by And~XI, XII, XIII, and MGC1, but also two less significant detections, with $S=3.9$ and 3.5. The CMDs of stars within $2'$ of these two detections are shown in the left-most panels of Figure~\ref{AndXI-XIII_detection_CMDs}, along with the CMDs of field annuli at a larger radius and with an identical coverage. If the least significant of the two detections, just above the detection threshold, shows no clear sign of an RGB at the distance of M31, the more significant detection does display a group of less than 10 stars which align themselves in CM space to produce a believable RGB that is not present in the field CMD. Of course, this is the very reason for the higher significance of this region of the sky and should not be used as a definite argument for the presence of a dwarf galaxy at this location. It does, however, bolster confidence in the results of the algorithm and its ability to isolate potential sparse dwarf galaxies. This particular detection is being followed up as a strong dwarf galaxy candidate.

\subsection{Global results}

\begin{figure*}
\centering
\includegraphics[width=0.8\hsize,angle=270]{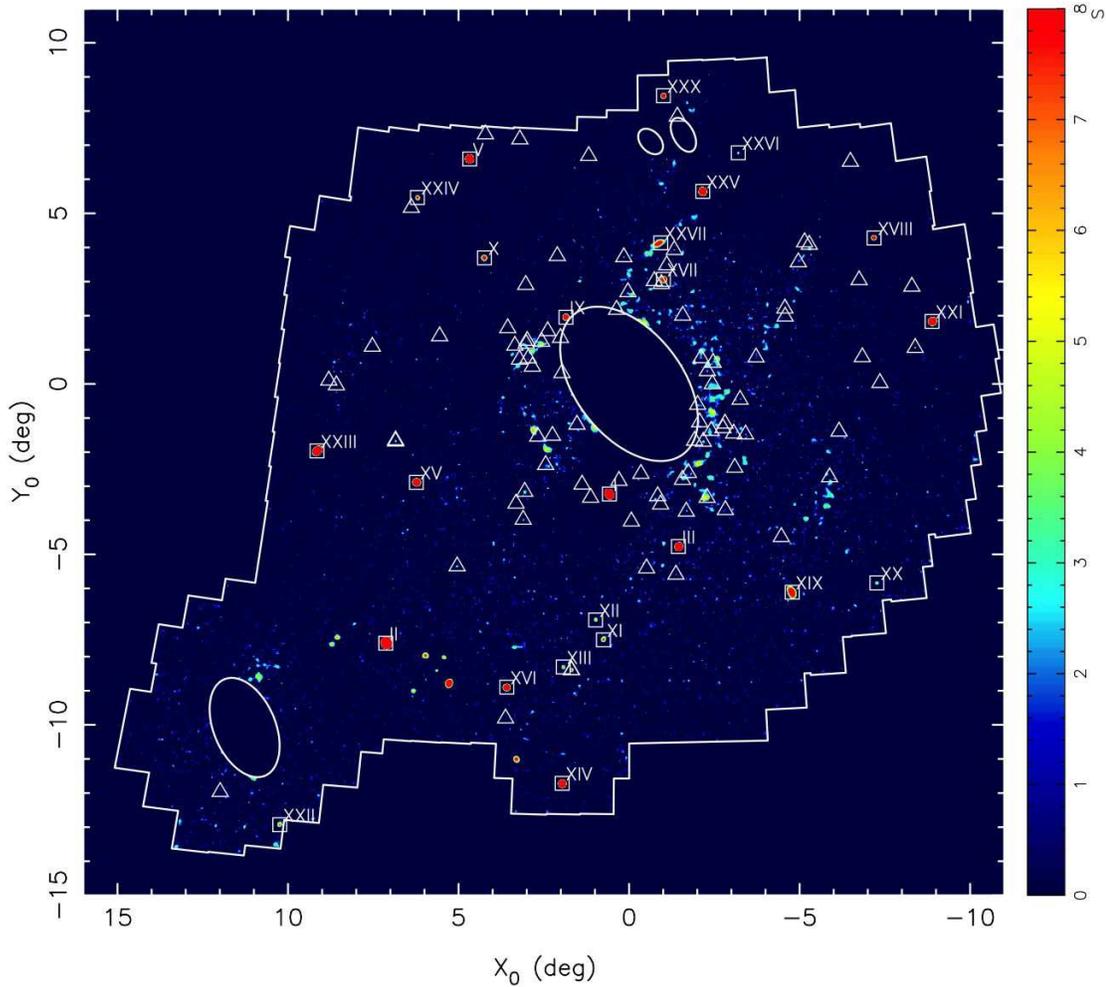}
\caption{\label{total_signal} Map of significance $S$ over the full PAndAS footprint, represented by the white polygon. The four large ellipses correspond to the regions around M31, M33, NGC~147 and NGC~185 that have been masked due to their high stellar density. The M31 and M33 ellipses have major axis lengths of $2.6\deg$ and $1.53\deg$, respectively, and an ellipticity of 0.4 in both cases. White squares represent all known dwarf galaxies within the PAndAS footprint, labeled by their Andromedan number, and white triangles represent known GCs. For both dwarf galaxies and globular clusters, systems that fall within the masked out regions are not shown. The mainly low signal value of the map when compared to the very structured nature of the M31 halo (Figure~\ref{M31_PAndAS}) is testimony to the ability of the algorithm to account for M31 halo contamination. All known dwarf galaxies are detected at high significance.}
\end{figure*}

Figure~\ref{total_signal} displays the signal map that results from the application of the algorithm to the full PAndAS survey. Most streams go unnoticed compared to the raw PAndAS maps. Comparing the global features of this map with the map of stellar structures (e.g. Figure~\ref{M31_PAndAS} or the bottom panel of Figure~\ref{contamination_data_model_comparison}) highlights the overall efficiency of the algorithm to overcome the structured nature of the M31 stellar halo in order to focus on very localized overdensities. Our simple model of the M31 contamination (a single broad RGB in CM space, distributed uniformly on the local sky) yields good results, except in complex regions of overlapping or clumpy streams for which the algorithm becomes less efficient (e.g. the inner halo regions to the east of M31, the overlapping streams C and D, the SW cloud, or the stellar debris around M33). There is no easy workaround that could deal with this issue and we'll tackle it through the choices of $S_\mathrm{th}$ over which detections are deemed significant.

The other main feature of this map is the high significance of all known dwarf galaxies (represented by white squares), along with the detections of some known GCs (represented by white triangles). As was the case for the regions around And~XI-XIII, there are also numerous intriguing outer halo, isolated detections that do not overlap with any known structure. Rather disappointingly for the search of dwarf galaxies, the grouping of very significant detections in the region with $5\deg\simlt X_0\simlt9\deg$ and $-9\deg\simlt Y_0\simlt-7\deg$ is produced by the NGC~507 group of background elliptical galaxies whose distance means that their GCs are identified as point sources in PAndAS. In addition, these GCs distribute themselves in CM space in a manner that is very similar to RGB stars from an M31 dwarf galaxy.

\begin{figure}
\centering
\includegraphics[width=0.95\hsize]{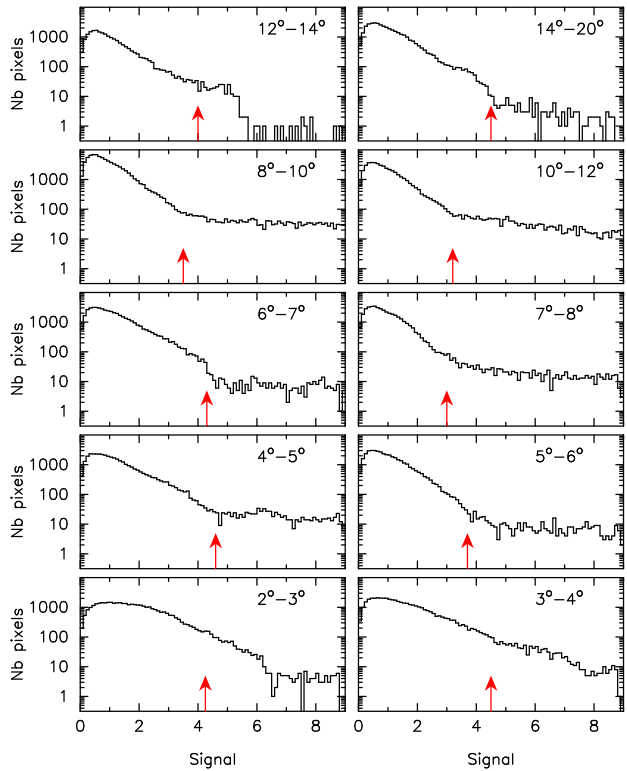}
\caption{\label{significance_hist_total} Significance distributions for circular annuli at different distances from M31, and the corresponding significance thresholds used for the detection selections in the different annuli (red arrows).}
\end{figure}

Significant detections are selected as above by enforcing a significance threshold. However, given the varying noise in Figure~\ref{total_signal}, we enforce thresholds that vary with distance from M31. This allows us to optimize the search for locally significant detections in outer halo regions and is highlighted by the change in shape of the histograms of $S$ values for different radial annuli, as shown in Figure~\ref{significance_hist_total}. In the inner regions of the survey, these histograms display a more extended distribution at the high end, whereas the histograms for large distances from M31 are more akin to that of the And~XI-XIII region (Figure~\ref{significance_hist_AndXI-XIII}), until they contain the stellar debris around M33 ($>12\deg$). The values of $S_\mathrm{th}$ are chosen for each annulus where there is a break between the bulk of the distribution and the high end, flat tail produced by detections of dwarf-galaxy-like detections. For the two small inner-most annuli ($<4\deg$), conservative values are selected before the break in the distributions in order to keep detection pixels at the cost of more false-positives.

\begin{figure*}
\centering
\includegraphics[width=0.8\hsize,angle=270]{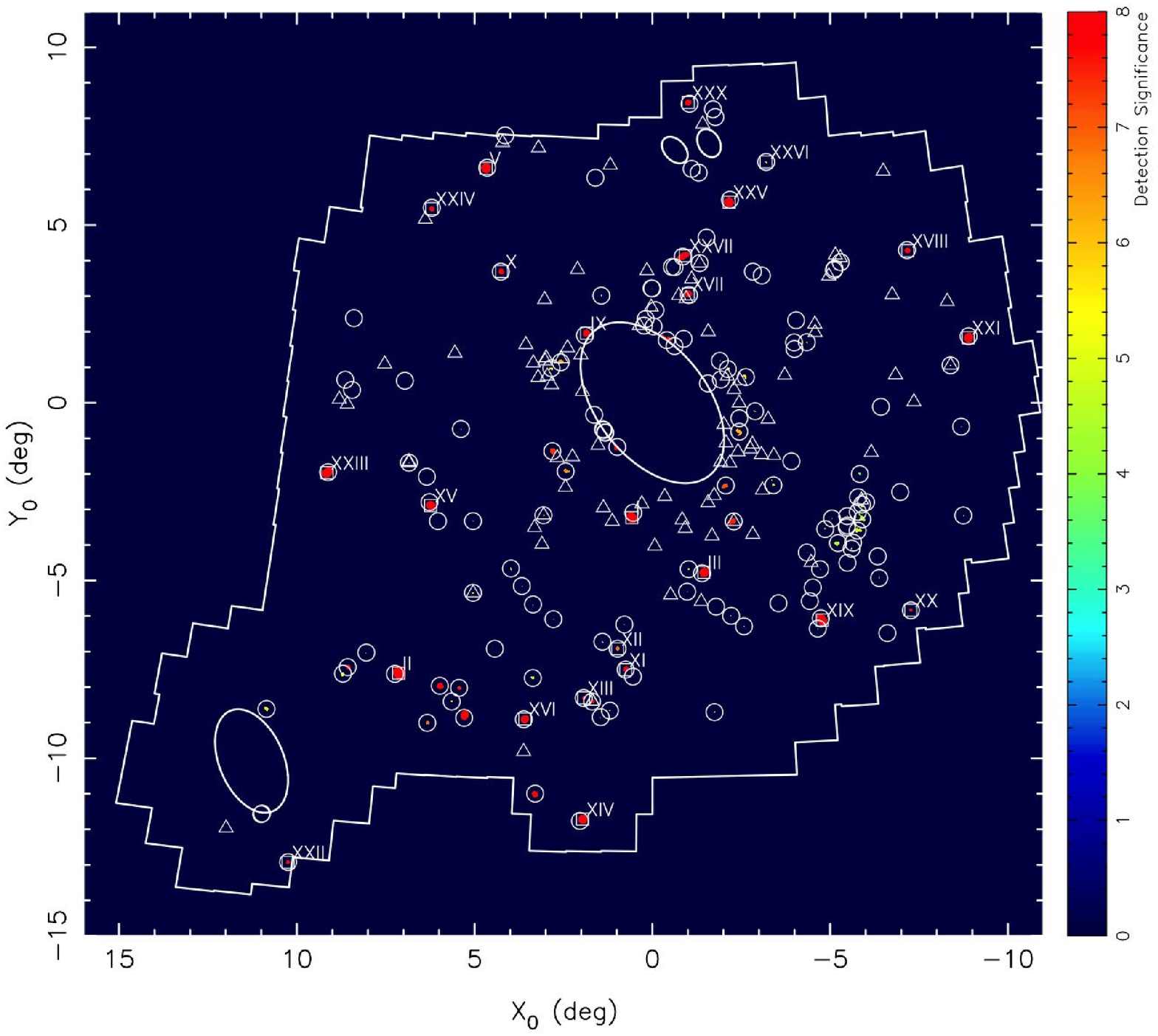}
\caption{\label{total_detections} Map of all detections found by the algorithm within the PAndAS footprint, delimited by the white polygon. As before, white squares and triangles represent known dwarf galaxies and GCs, respectively. The detections are circled in white and a detection's pixels are color-coded by its significance.}
\end{figure*}

Applying these significance thresholds yields the map of detections presented in Figure~\ref{total_detections}. All of these detections are listed in Table~1, along with their nature, when known, or comments on their location when they overlap with known structures. The code saturates at $S = 8.5$ for numerical reasons so the first detections are given in no particular order. Dwarf galaxies dominate the top of the list, as can be expected from their direct discovery, and the least significant, And~XXVI, is still found with $S=5.9$.

The original PAndAS images were inspected at the location of all 143~detections. This inspection confirmed the detections due to the NGC~507 group of background galaxies, allowed for the recognition of some significantly detected known GCs and ECs (detections~22, 29, 43, 51, 69, 70, and 72) as well as one new EC, and point out the odd artefacts (detection~112 and 129). As could be expected from the signal map, a large fraction of significant detections overlap with known stellar structures. At this stage, it is not possible to disentangle real stellar clumps in the streams from artificial detections that could be due to a locally inadequate M31 halo contamination model. We therefore chose to provide an unbiased list that includes the raw output of the algorithm.

\subsubsection{Discovery of the EC PAndAS-31}
\label{section:newEC}
The algorithm is able to find some GCs when these are bright or extended enough to be resolved as compact overdensities of RGB stars at the distance of M31, as exemplified by the detection of MGC1. Eight such clusters are found automatically: Bol~514 and 289 \citep{galleti04} in the inner regions, along with the outer halo cluster H27/MCGC10 \citep{huxor08,mackey07} and HEC12/EC4 \citep{huxor08,mackey06,collins09a}, the odd system PAndAS-48 \citep{mackey13}, and three other outer halo GCs that will be presented in detail in Huxor et al. (2013, in preparation; PAndAS-02, PAndAS-26, and PAndAS-49). In addition, the algorithm was also able to discover a new EC, PAndAS-31 (detection~30), located at $(\alpha, \delta) = $(00:39:54.7, +43:02:59), and undetected by the thorough search of all PAndAS images by Huxor et al. (in prep.).

\begin{figure}
\centering
\includegraphics[width=0.9\hsize]{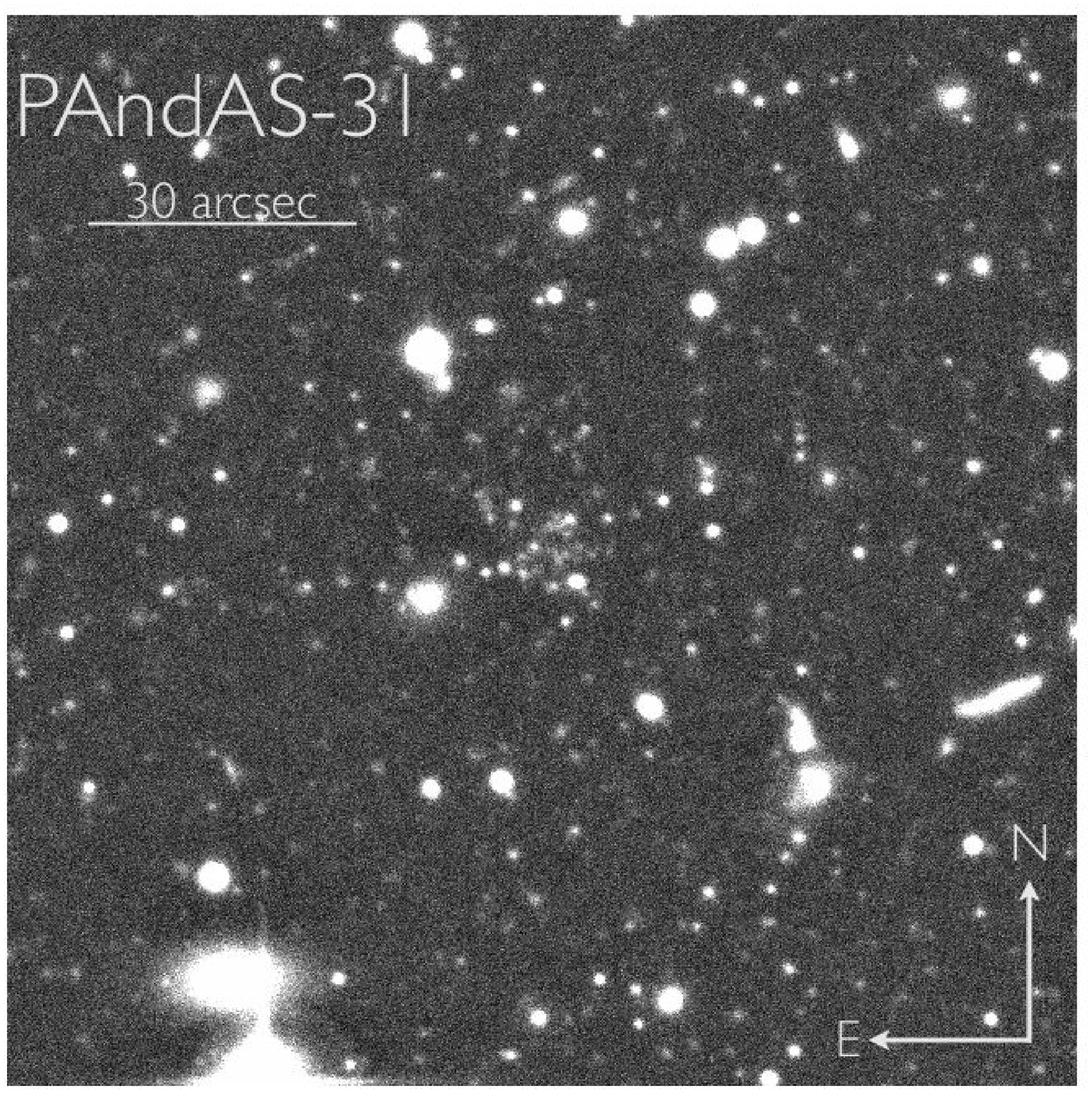}
\caption{\label{PAndAS-31} PAndAS stacked image centered on the EC PAndAS-31 detected by the algorithm. The image is in the $i$ band and has a size of $2'\times2'$. North is to the top and east is to the left.}
\end{figure}

PAndAS-31 is a faint EC (see Figure~\ref{PAndAS-31}) that went unnoticed in the high stellar density region of the M31 inner halo. Its total magnitude ($M_V = -4.4\pm0.2$) and size ($r_h \simeq20\pc$) are typical of low luminosity ECs (e.g. \citealt{huxor11}). For these properties, the search of Huxor et al. (in preparation) is only $\sim50\%$ complete, whereas the search algorithm triggers on the few bright RGB stars that dominate the light of the cluster.

The full list of properties for PAndAS-31 will be given in the main PAndAS GC catalog (Huxor et al., in preparation).

\subsubsection{Isolated detections}

The visual inspection of the images of a handful of the isolated detections, along with their CMD and stellar distribution, lead us to believe they are likely genuine dwarf galaxies, fainter than the currently known ones ($M_V\simgt-6.5$). They share the properties of the most significant detection in the region of And~XI-XIII, presented in Figure~\ref{AndXI-XIII_detection_CMDs}. Yet, the low density of most of the detections listed in Table~1 raises the question of their reality as the PAndAS data alone cannot clearly confirm they are authentic. Deep photometric follow-up is necessary for this confirmation. It is however possible to check the collective validity of the detections by stacking them.

\begin{figure*}
\centering
\includegraphics[width=0.4\hsize,angle=270]{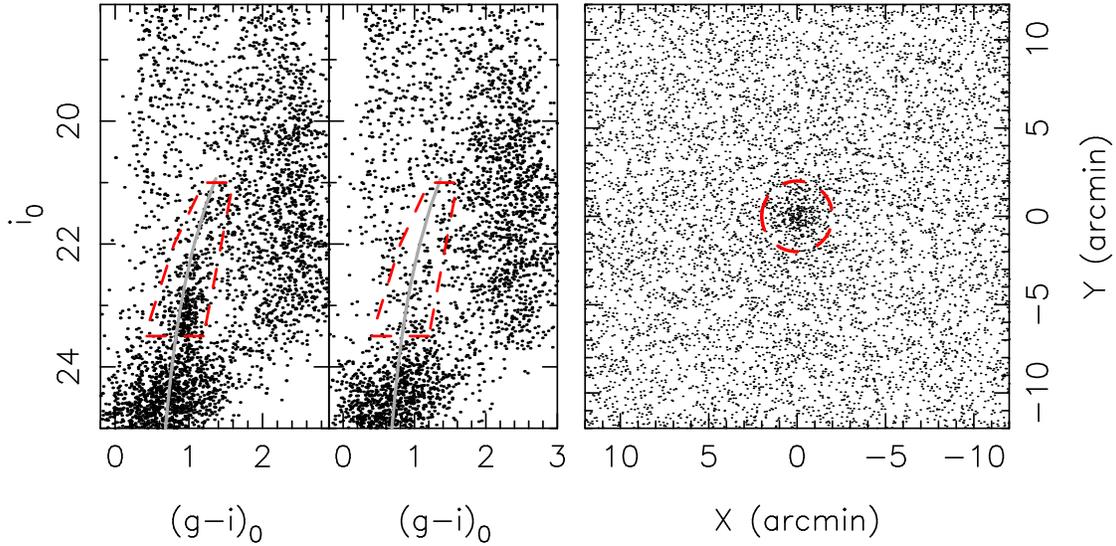}
\caption{\label{combined_CMDs} Combined properties of all 39~isolated detections of Table~1. \emph{Left-hand panel:} CMD of all point sources within a radius of $2'$ of the centroids of all the detections. An RGB-like grouping of stars is visible near the fiducial old M31 stellar population with $\FeH=-1.7$ (gray line). The red dashed polygon is used to isolate RGB-like stars. \emph{Middle panel:} Field CMD of all stars within an annulus of identical coverage located at $15'$ from the detections' centroids. \emph{Right-hand panel:} The stacked distribution of RGB-like point sources around the isolated 39~detections. A clear overdensity is visible in the red dashed circle that corresponds to a radius of $2'$.}
\end{figure*}

\begin{figure}
\includegraphics[width=0.75\hsize,angle=270]{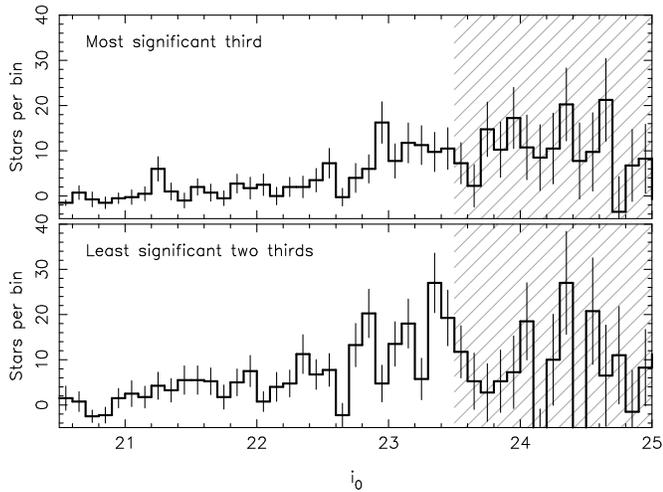}
\caption{\label{combined_LFs} Contamination corrected LFs of the third most significant isolated detections listed in table~\ref{table:detection_list} (top panel) and the remain two thirds (bottom panel). The error bars were calculated assuming Poisson statistics. The region hashed in gray corresponds to stars that were not used by the search algorithm to determine the significance of the detections. The non-zero counts in the hashed region of the top panel suggests these detections include genuine dwarf galaxies.}
\end{figure}

As can be seen on Figure~\ref{combined_CMDs}, the collective CMD of all catalog sources within $2'$ of the 39~isolated detections contains a clear grouping of RGB-like stars at $i_0\simlt22.0$ and $(g-i)_0\sim1.0$. At the same time, the stellar distribution of these stars with respect to the centroids of the detections displays a significant stellar overdensity of a few arcminutes on the side, typical of faint M31 dwarf galaxies. Alone, the presence of these features is not a proof of dwarf-galaxy nature of these detections, but only confirms that the algorithm does find what it is tailored to find (spatially clumped overdensities of stars that align themselves in the CMD). However, it is interesting to note that the background-subtracted luminosity function (LF) of all stars within $2'$ of these detections remains overdense below the magnitude limit used to select stars that were used by the search algorithm, in particular for the most significant of the isolated detections. This is shown in Figure~\ref{combined_LFs} for which we have split the sample between the third most significant detections (top panel), and the least significant two thirds (bottom panel). These LFs are built from stars with RGB-like colors (accounting for a widening color range as photometric uncertainties increase) that are located within $2'$ of a detections' center. They are further corrected from contamination by subtracting the LF of a corresponding field population, scaled to the same radius. Looking at this figure, one should keep in mind that the data become increasingly incomplete below $i_0=23.5$, which plays to counteract the natural increase of the LF at fainter magnitudes. One can nevertheless notice significant non-zero counts for the LF below this limit for the top third most significant isolated detections, confirming the present of stars that aren't accounted for in the field. These stars were not used by the search algorithm, thereby yielding an independent detections of the reality of at least some of these detections. The absence of a clear bump in the LF that would correspond to the presence of a well-defined horizontal branch could easily be explained by the likely different distances to all these detections, blurring out any signal from a faint horizontal branch. The LF of the least significant isolated detections listed in Table~\ref{table:detection_list} (bottom panel of Figure~\ref{combined_LFs}) shows a very marginal detection of extra stars fainter than $i_0=23.5$, making the reality of these detection less likely, although, if they are real, these would likely be very faint dwarf galaxies, and correspondingly much more difficult to find without deep and complete data.

As a final note, it is worth mentioning that our detection~124 overlaps with one of the \textsc{Hi} clouds discovered by \citet[their detection 4]{wolfe13}.

\section{Conclusions}
We have presented a generic algorithm to search for stellar overdensities in a photometric catalog and applied this algorithm to trawl the PAndAS data set of the surroundings of the Andromeda galaxy for dwarf galaxies. All known dwarf galaxies have been found at high significance ($S>5.9$) in the list of 143 most significant detections returned by the algorithm. Some GCs are also detected, and a new EC was discovered (PAndAS-31). At least some of the algorithm's unknown detections share the properties of dwarf galaxies fainter than those currently known around M31. They should be followed up by the community with deep photometry, with the hope of confirming the first very faint M31 satellite dwarf galaxies, likely with properties similar to Leo~IV, or UMa~II around the MW, but without the strong distance bias in the search for such systems within the SDSS or the Pan-STARRS1 survey \citep{koposov08,walsh09}.

The development of an efficient dwarf-galaxy-search algorithm is not only necessary to potentially unveil even fainter dwarf galaxies than those currently known (the goal of this paper), but also it is mandatory to derive the dwarf galaxy completeness limits of the PAndAS survey. These are cardinal in understanding the global properties of the satellite system (structure of the satellite system, (an)isotropy, luminosity function, etc) as they enable a reliable correction from the survey's observational effects, as well as the strong variation of the MW and M31 stellar halo contamination throughout the survey's footprint. They will be derived in an upcoming contribution.

Finally, we wish to emphasize that the search algorithm presented here is applicable to the search of any source distribution in a photometric catalog and, even more generically, to \emph{any} source distribution in \emph{any} catalog, inasmuch as reliable models of the signal and contamination can be computed.

\acknowledgments
We are grateful to the CFHT observing team for gathering the PAndAS images, and for their continued support throughout the project. N.F.M. thanks Hans-Walter Rix and David Hogg for numerous conversations, essential to the development of this work. G.F.L. gratefully acknowledges financial support for his ARC Future Fellowship (FT100100268) and through the award of an ARC Discovery Project (DP110100678). This work is based on observations obtained with MegaPrime/MegaCam, a joint project of CFHT and CEA/DAPNIA, at the Canada-France-Hawaii Telescope, which is operated by the National Research Council (NRC) of Canada, the Institut National des Sciences de l'Univers of the Centre National de la Recherche Scientifique (CNRS) of France, and the University of Hawaii.


\clearpage

\begin{center}
\begin{longtable*}{l|ccccccl}
\caption{\label{table:detection_list}List of significant detections.} \\

Detection & $\alpha$ (J2000) & $\delta$ (J2000) & $X_0$ (deg) & $Y_0$ (deg) & $S$ & $S_\mathrm{th}$ & Remark\\ \hline 
\endfirsthead

\multicolumn{8}{c}
{\tablename\ \thetable\ -- \textit{Continued from previous page}} \\
Detection & $\alpha$ (J2000) & $\delta$ (J2000) & $X_0$ (deg) & $Y_0$ (deg) & $S$ & $S_\mathrm{th}$ & Remark \\
\hline
\endhead

Detection   1 &  00:19:45 & +35:05:36 & $-4.7$ & $-6.1$ & $>8.5$ & 3.5 & And~XIX\\
Detection   2 &  00:35:49 & +36:28:20 & $-1.4$ & $-4.8$ & $>8.5$ & 3.7 & And~III\\
Detection   3 &  01:10:07 & +47:38:55 &  $+4.6$ &  $+6.6$ & $>8.5$ & 3.2 & And~V\\
Detection   4 &  01:17:00 & +33:23:19 &  $+7.2$ & $-7.6$ & $>8.5$ & 4.0 & And~II\\
Detection   5 &  01:07:27 & +32:19:31 &  $+5.3$ & $-8.9$ & $>8.5$ & 4.0 & GC system of background galaxies\\
 & & & & & & & (NGC~382, 383, 384, 385, 386)\\
Detection   6 &  23:54:43 & +42:30:28 & $-8.9$ &  $+1.9$ & $>8.5$ & 3.2 & And~XXI\\
Detection   7 &  00:30:03 & +46:55:09 & $-2.2$ &  $+5.7$ & $>8.5$ & 3.0 & And~XXV\\
Detection   8 &  01:29:13 & +38:44:08 &  $+9.1$ & $-2.0$ & $>8.5$ & 3.2 & And~XXIII\\
Detection   9 &  00:36:21 & +49:36:40 & $-1.0$ &  $+8.4$ & $>8.5$ & 3.2 & Cas~II/And~XXX\\
Detection  10 &  00:51:55 & +29:38:53 &  $+2.0$ & $-11.8$ & $>8.5$ & 4.0 & And~XIV\\
Detection  11 &  00:45:28 & +38:10:28 &  $+0.5$ & $-3.1$ & $>8.5$ & 4.6 & And~I\\
Detection  12 &  00:59:38 & +32:21:55 &  $+3.6$ & $-8.9$ & $>8.5$ & 3.2 & And~XVI\\
Detection  13 &  00:37:53 & +45:22:06 & $-0.9$ &  $+4.1$ & $>8.5$ & 3.7 & And~XXVII\\
Detection  14 &  00:36:59 & +44:17:12 & $-1.0$ &  $+3.0$ & $>8.5$ & 4.6 & And~XVII\\
Detection  15 &  01:06:42 & +44:47:42 &  $+4.3$ &  $+3.7$ & $>8.5$ & 4.3 & And~X\\
Detection  16 &  01:14:32 & +38:12:23 &  $+6.3$ & $-2.8$ & $>8.5$ & 3.0 & And~XV\\
Detection  17 &  00:02:22 & +45:06:37 & $-7.2$ &  $+4.3$ & $>8.5$ & 3.2 & And~XVIII\\
Detection  18 &  00:53:07 & +43:08:06 &  $+1.9$ &  $+1.9$ & $>8.5$ & 4.5 & And~IX\\
Detection  19 &  01:18:30 & +46:23:26 &  $+6.2$ &  $+5.5$ & $>8.5$ & 3.2 & And~XXIV\\
Detection  20 &  00:57:44 & +30:21:07 &  $+3.3$ & $-11.0$ & $>8.5$ & 4.0 & GC system of background galaxy (NGC~315)\\
Detection  21 &  00:46:20 & +33:48:16 &  $+0.8$ & $-7.5$ & $>8.5$ & 3.5 & And~XI\\
Detection  22 &  00:50:44 & +32:54:31 &  $+1.7$ & $-8.4$ & $>8.5$ & 3.2 & MGC1 \citep{martin06b}\\
Detection  23 &  01:10:57 & +33:09:28 &  $+6.0$ & $-8.0$ & $>8.5$ & 3.2 & GC system of background galaxy (NGC~410)\\
Detection  24 &  01:27:40 & +28:05:53 & $+10.2$ & $-12.9$ & $>8.5$ & 4.5 & And~XII\\
Detection  25 &  00:51:51 & +33:00:07 &  $+1.9$ & $-8.3$ & 8.5 & 3.2 & And~XIII\\
Detection  26 &  00:47:54 & +40:00:59 &  $+1.0$ & $-1.2$ & 7.8 & 4.2 & overlap with M31 inner halo stellar structure\\
Detection  27 &  00:07:30 & +35:07:52 & $-7.3$ & $-5.8$ & 7.8 & 3.2 & And~XX\\
Detection  28 &  01:23:16 & +33:27:28 &  $+8.6$ & $-7.4$ & 7.7 & 4.0 & GC system of background galaxies\\
 & & & & & & & (NGC~495 and 499)\\
Detection  29 &  00:31:10 & +37:54:00 & $-2.3$ & $-3.3$ & 7.5 & 3.7 & GC Bol 514 (H6, MCGC4)\\
 & & & & & & & \citep{galleti05,huxor08,mackey07}\\
Detection  30 &  00:40:32 & +43:01:48 & $-0.4$ &  $+1.8$ & 7.5 & 4.2 & overlap with M31 inner halo stellar structure\\
 & & & & & & & includes new EC PAndAS-31, see \S~\ref{section:newEC}\\
Detection  31 &  01:08:24 & +33:08:07 &  $+5.4$ & $-8.0$ & 7.5 & 3.2 & GC system of background galaxies\\
 & & & & & & & (NGC~392, 394, 397)\\
Detection  32 &  00:57:23 & +39:50:58 &  $+2.8$ & $-1.4$ & 7.4 & 4.6 & overlap with minor axis stream D \citep{ibata07}\\
Detection  33 &  00:32:09 & +38:54:41 & $-2.1$ & $-2.3$ & 7.1 & 4.6 & overlap with M31 inner halo stellar structure\\
Detection  34 &  01:12:11 & +32:07:28 &  $+6.3$ & $-9.0$ & 6.9 & 4.0 & GC system of background galaxy (NGC~420)\\
Detection  35 &  00:47:27 & +34:22:34 &  $+1.0$ & $-6.9$ & 6.9 & 3.0 & And~XII\\
Detection  36 &  00:55:20 & +39:17:54 &  $+2.4$ & $-1.9$ & 6.6 & 4.6 & overlap with minor axis stream D\\
Detection  37 &  00:56:41 & +42:21:13 &  $+2.6$ &  $+1.1$ & 6.3 & 4.5 & overlap with minor axis stream C/D \citep{ibata07}\\
Detection  38 &  00:29:54 & +40:24:45 & $-2.4$ & $-0.8$ & 6.2 & 4.5 & overlap with M31 inner halo stellar structure\\
Detection  39 &  00:23:45 & +47:54:40 & $-3.2$ &  $+6.8$ & 5.9 & 3.5 & And~XXVI\\
Detection  40 &  00:28:36 & +41:56:09 & $-2.6$ &  $+0.7$ & 5.7 & 4.5 & overlap with M31 inner halo stellar structure\\
Detection  41 &  01:09:14 & +32:44:59 &  $+5.6$ & $-8.4$ & 5.7 & 4.0 & GC system of background galaxy (NGC~403) \\
Detection  42 &  01:32:57 & +32:05:37 & $+10.9$ & $-8.6$ & 5.7 & 4.5 & overlap with M33 stellar debris\\
Detection  43 &  23:57:57 & +41:46:32 & $-8.4$ &  $+1.1$ & 5.5 & 3.2 & GC PAndAS-02 (Huxor et al., in preparation)\\
Detection  44 &  00:58:03 & +42:10:35 &  $+2.8$ &  $+1.0$ & 5.3 & 4.5 & overlap with minor axis stream C/D\\
Detection  45 &  01:23:46 & +33:16:05 &  $+8.7$ & $-7.6$ & 5.3 & 4.0 & GC system of background galaxies\\
 & & & & & & & (NGC~507, 508)\\
Detection  46 &  00:31:18 & +42:10:46 & $-2.1$ &  $+0.9$ & 5.2 & 4.5 & overlap with M31 inner halo stellar structure\\
Detection  47 &  00:50:45 & +44:15:41 &  $+1.4$ &  $+3.0$ & 5.2 & 4.6 & Andromeda NE \citep{zucker04a}\\
Detection  48 &  01:07:28 & +35:46:49 &  $+5.0$ & $-5.3$ & 5.2 & 3.5 & GC H27/MCGC10 \citep{huxor08,mackey07}\\
Detection  49 &  00:49:57 & +40:29:04 &  $+1.4$ & $-0.8$ & 5.2 & 4.2 & overlap with M31 inner halo stellar structure\\
Detection  50 &  00:39:21 & +42:51:42 & $-0.6$ &  $+1.6$ & 5.0 & 4.2 & overlap with M31 inner halo stellar structure\\
Detection  51 &  00:58:16 & +38:02:46 &  $+3.1$ & $-3.2$ & 5.0 & 3.7 & HEC12 (MCEC4)\\
 & & & & & & &  \citep{huxor08,mackey06,collins09a}\\
Detection  52 &  00:32:28 & +42:25:11 & $-1.9$ &  $+1.2$ & 4.9 & 4.5 & overlap with M31 inner halo stellar structure\\
Detection  53 &  00:13:47 & +37:27:24 & $-5.8$ & $-3.6$ & 4.9 & 3.0 & overlap with SW cloud\\
Detection  54 &  01:10:57 & +40:19:01 &  $+5.4$ & $-0.7$ & 4.9 & 4.3 & \\
Detection  55 &  00:49:43 & +40:25:37 &  $+1.3$ & $-0.8$ & 4.9 & 4.2 & overlap with M31 inner halo stellar structure\\
Detection  56 &  00:27:27 & +40:58:07 & $-2.9$ & $-0.2$ & 4.9 & 4.5 & overlap with NW stream\\
Detection  57 &  00:42:46 & +44:28:42 &  $+0.0$ &  $+3.2$ & 4.8 & 4.6 & overlap with And~XXVII stream\\
Detection  58 &  00:42:15 & +43:52:18 & $-0.1$ &  $+2.6$ & 4.8 & 4.5 & overlap with M31 inner halo stellar structure\\
Detection  59 &  00:39:14 & +45:04:22 & $-0.6$ &  $+3.8$ & 4.8 & 4.6 & overlap with And~XXVII stream\\
Detection  60 &  00:49:41 & +40:24:38 &  $+1.3$ & $-0.8$ & 4.7 & 4.2 & overlap with M31 inner halo stellar structure\\
Detection  61 &  00:32:22 & +41:53:39 & $-1.9$ &  $+0.7$ & 4.6 & 4.5 & overlap with M31 inner halo stellar structure\\
Detection  62 &  01:31:38 & +29:18:15 & $+11.0$ & $-11.6$ & 4.6 & 4.5 & overlap with M33 stellar debris\\
Detection  63 &  00:50:02 & +40:30:55 &  $+1.4$ & $-0.7$ & 4.6 & 4.2 & overlap with M31 inner halo stellar structure\\
Detection  64 &  00:39:39 & +45:04:55 & $-0.5$ &  $+3.8$ & 4.6 & 4.6 & overlap with And~XXVII stream\\
Detection  65 &  00:44:03 & +43:26:19 &  $+0.2$ &  $+2.2$ & 4.6 & 4.5 & overlap with M31 inner halo stellar structure\\
Detection  66 &  00:42:32 & +43:24:51 & $-0.0$ &  $+2.1$ & 4.5 & 4.5 & overlap with M31 inner halo stellar structure\\
Detection  67 &  00:29:49 & +40:47:12 & $-2.4$ & $-0.4$ & 4.5 & 4.5 & overlap with M31 inner halo stellar structure\\
Detection  68 &  00:43:46 & +43:37:19 &  $+0.2$ &  $+2.4$ & 4.5 & 4.5 & overlap with M31 inner halo stellar structure\\
Detection  69 &  00:34:21 & +41:47:46 & $-1.6$ &  $+0.5$ & 4.5 & 4.2 & GC Bol~289 \citep{galleti04}\\
Detection  70 &  01:17:58 & +39:14:56 &  $+6.8$ & $-1.7$ & 4.5 & 3.5 & double GCs PAndAS-53 and PAndAS-54\\
 & & & & & & & (Huxor et al., in preparation)\\
Detection  71 &  00:37:40 & +36:35:38 & $-1.0$ & $-4.7$ & 4.4 & 3.7 & \\
Detection  72 &  00:35:13 & +45:10:37 & $-1.3$ &  $+3.9$ & 4.4 & 3.7 & GC PAndAS-27 (Huxor et al., in preparation)\\
Detection  73 &  00:37:55 & +35:57:57 & $-1.0$ & $-5.3$ & 4.4 & 4.3 & \\
Detection  74 &  00:16:39 & +37:09:02 & $-5.2$ & $-3.9$ & 4.4 & 3.0 & overlap with SW cloud\\
Detection  75 &  00:25:17 & +38:52:06 & $-3.4$ & $-2.3$ & 4.3 & 3.7 & \\
Detection  76 &  00:12:51 & +38:10:17 & $-5.9$ & $-2.9$ & 4.3 & 3.0 & overlap with SW cloud\\
Detection  77 &  00:51:24 & +40:54:41 &  $+1.6$ & $-0.3$ & 4.3 & 4.2 & overlap with M31 inner halo stellar structure\\
Detection  78 &  00:37:56 & +43:03:30 & $-0.9$ &  $+1.8$ & 4.3 & 4.2 & overlap with M31 inner halo stellar structure\\
Detection  79 &  00:12:59 & +37:45:33 & $-5.9$ & $-3.3$ & 4.3 & 3.0 & overlap with SW cloud\\
Detection  80 &  00:58:43 & +33:30:22 &  $+3.4$ & $-7.7$ & 4.2 & 3.2 & \\
Detection  81 &  00:12:54 & +44:58:26 & $-5.3$ &  $+4.0$ & 4.2 & 3.0 & overlap with NW stream\\
Detection  82 &  00:34:33 & +32:36:52 & $-1.7$ & $-8.7$ & 4.1 & 3.2 & \\
Detection  83 &  01:21:02 & +33:54:07 &  $+8.0$ & $-7.0$ & 4.1 & 4.0 & GC system of background galaxy (UGC~878)\\
Detection  84 &  00:19:05 & +42:49:13 & $-4.3$ &  $+1.7$ & 4.1 & 3.7 & overlap with NW stream\\
Detection  85 &  00:25:26 & +44:45:43 & $-3.1$ &  $+3.6$ & 4.0 & 3.7 & \\
Detection  86 &  00:22:29 & +39:30:50 & $-3.9$ & $-1.6$ & 4.0 & 3.7 & \\
Detection  87 &  00:34:02 & +45:53:03 & $-1.5$ &  $+4.6$ & 3.9 & 3.7 & overlap with NW stream\\
Detection  88 &  00:48:21 & +32:39:51 &  $+1.2$ & $-8.7$ & 3.9 & 3.2 & see Figure~\ref{AndXI-XIII_detection_CMDs} (top panels)\\
Detection  89 &  00:13:18 & +38:23:10 & $-5.8$ & $-2.7$ & 3.9 & 3.0 & overlap with SW cloud\\
Detection  90 &  00:12:46 & +39:01:31 & $-5.8$ & $-2.0$ & 3.8 & 3.0 & overlap with SW cloud\\
Detection  91 &  00:20:14 & +34:48:40 & $-4.7$ & $-6.4$ & 3.8 & 3.5 & \\
Detection  92 &  23:58:54 & +37:35:26 & $-8.7$ & $-3.2$ & 3.8 & 3.2 & \\
Detection  93 &  01:03:59 & +34:16:01 & $+4.4$ & $-6.9$ & 3.8 & 3.2 & \\
Detection  94 &  00:26:48 & +44:52:27 & $-2.8$ &  $+3.7$ & 3.8 & 3.7 & \\
Detection  95 &  00:31:58 & +35:15:55 & $-2.2$ & $-6.0$ & 3.7 & 3.0 & \\
Detection  96 &  00:11:18 & +36:41:50 & $-6.3$ & $-4.3$ & 3.7 & 3.5 & \\
Detection  97 &  00:21:01 & +42:50:10 & $-4.0$ &  $+1.7$ & 3.7 & 3.7 & overlap with NW stream\\
Detection  98 &  00:20:29 & +43:27:14 & $-4.0$ &  $+2.3$ & 3.7 & 3.7 & \\
Detection  99 &  01:02:27 & +36:30:56 & $+4.0$ & $-4.7$ & 3.7 & 3.0 & overlap with stream B \citep{ibata07}\\
Detection 100 &  00:11:22 & +36:05:44 & $-6.4$ & $-4.9$ & 3.7 & 3.2 & \\
Detection 101 &  00:14:49 & +36:58:02 & $-5.6$ & $-4.1$ & 3.7 & 3.0 & overlap with SW cloud\\
Detection 102 &  00:13:55 & +44:48:27 & $-5.1$ &  $+3.8$ & 3.7 & 3.0 & overlap with NW stream\\
Detection 103 &  00:35:02 & +47:41:48 & $-1.3$ &  $+6.5$ & 3.7 & 3.0 & overlap with NGC~147 stream\\
Detection 104 &  00:15:04 & +37:52:55 & $-5.5$ & $-3.2$ & 3.6 & 3.0 & overlap with SW cloud\\
Detection 105 &  00:21:05 & +35:35:11 & $-4.4$ & $-5.6$ & 3.6 & 3.5 & overlap with SW cloud\\
Detection 106 &  00:18:17 & +37:33:42 & $-4.9$ & $-3.6$ & 3.6 & 3.0 & \\
Detection 107 &  00:07:18 & +38:25:23 & $-7.0$ & $-2.5$ & 3.5 & 3.5 & \\
Detection 108 &  00:15:33 & +36:34:51 & $-5.5$ & $-4.5$ & 3.5 & 3.5 & overlap with SW cloud\\
Detection 109 &  00:31:59 & +49:13:25 & $-1.8$ &  $+8.0$ & 3.5 & 3.2 & overlap with NGC~147 stream\\
Detection 110 &  00:45:20 & +33:36:34 & $+0.5$ & $-7.7$ & 3.5 & 3.5 & see Figure~\ref{AndXI-XIII_detection_CMDs} (bottom panels)\\
Detection 111 &  23:57:34 & +40:02:45 & $-8.7$ & $-0.7$ & 3.5 & 3.2 & \\
Detection 112 &  00:52:11 & +47:32:55 &  $+1.6$ &  $+6.3$ & 3.5 & 3.0 & star forming regions in background NGC~278\\
Detection 113 &  00:17:13 & +37:50:46 & $-5.1$ & $-3.3$ & 3.4 & 3.0 & \\
Detection 114 &  00:49:33 & +32:28:53 &  $+1.5$ & $-8.8$ & 3.4 & 3.2 & \\
Detection 115 &  00:10:56 & +34:33:30 & $-6.6$ & $-6.5$ & 3.3 & 3.2 & \\
Detection 116 &  00:36:12 & +47:48:29 & $-1.1$ &  $+6.6$ & 3.3 & 3.0 & overlap with NGC~147 stream\\
Detection 117 &  01:28:33 & +41:20:35 &  $+8.6$ &  $+0.6$ & 3.3 & 3.2 & \\
Detection 118 &  01:28:30 & +43:04:24 &  $+8.4$ &  $+2.4$ & 3.3 & 3.2 & \\
Detection 119 &  01:00:48 & +36:02:39 &  $+3.7$ & $-5.2$ & 3.3 & 3.0 & overlap with stream B\\
Detection 120 &  00:49:31 & +34:34:02 &  $+1.4$ & $-6.7$ & 3.3 & 3.0 & \\
Detection 121 &  00:21:05 & +36:56:28 & $-4.3$ & $-4.2$ & 3.3 & 3.0 & overlap with SW cloud\\
Detection 122 &  00:19:23 & +36:27:28 & $-4.7$ & $-4.7$ & 3.3 & 3.0 & overlap with SW cloud\\
Detection 123 &  01:27:22 & +41:05:23 &  $+8.4$ &  $+0.4$ & 3.3 & 3.2 & \\
Detection 124 &  01:08:11 & +37:46:21 &  $+5.0$ & $-3.3$ & 3.3 & 3.0 & overlap with detection~4 of \citet{wolfe13}\\
Detection 125 &  00:46:34 & +35:03:09 &  $+0.8$ & $-6.2$ & 3.3 & 3.0 & \\
Detection 126 &  00:59:08 & +35:32:51 &  $+3.4$ & $-5.7$ & 3.2 & 3.0 & overlap with stream B\\
Detection 127 &  00:32:22 & +49:26:17 & $-1.7$ &  $+8.3$ & 3.2 & 3.2 & overlap with NGC~147 stream\\
Detection 128 &  01:07:32 & +48:34:51 &  $+4.1$ &  $+7.5$ & 3.2 & 3.2 & at the edge of PAndAS\\
Detection 129 &  00:08:50 & +40:50:43 & $-6.4$ & $-0.1$ & 3.2 & 3.0 & bright star\\
Detection 130 &  00:12:14 & +38:13:41 & $-6.0$ & $-2.8$ & 3.1 & 3.0 & overlap with SW cloud\\
Detection 131 &  00:20:34 & +35:57:35 & $-4.5$ & $-5.2$ & 3.1 & 3.0 & overlap with SW cloud\\
Detection 132 &  00:33:57 & +35:31:51 & $-1.8$ & $-5.7$ & 3.1 & 3.0 & \\
Detection 133 &  01:15:15 & +38:54:29 &  $+6.3$ & $-2.1$ & 3.1 & 3.0 & \\
Detection 134 &  00:30:14 & +34:58:00 & $-2.6$ & $-6.3$ & 3.1 & 3.0 & \\
Detection 135 &  00:15:14 & +37:34:10 & $-5.5$ & $-3.5$ & 3.1 & 3.0 & overlap with SW cloud\\
Detection 136 &  01:19:49 & +41:30:57 &  $+7.0$ &  $+0.6$ & 3.1 & 3.0 & \\
Detection 137 &  00:14:05 & +44:46:07 & $-5.1$ &  $+3.7$ & 3.1 & 3.0 & \\
Detection 138 &  00:15:11 & +37:37:06 & $-5.5$ & $-3.5$ & 3.0 & 3.0 & overlap with SW cloud\\
Detection 139 &  00:56:18 & +35:09:26 &  $+2.8$ & $-6.1$ & 3.0 & 3.0 & \\
Detection 140 &  00:14:33 & +37:07:43 & $-5.6$ & $-3.9$ & 3.0 & 3.0 & \\
Detection 141 &  01:13:05 & +37:41:59 &  $+6.0$ & $-3.3$ & 3.0 & 3.0 & \\
Detection 142 &  00:13:31 & +37:58:29 & $-5.8$ & $-3.1$ & 3.0 & 3.0 & overlap with SW cloud\\
Detection 143 &  00:25:24 & +35:34:20 & $-3.5$ & $-5.6$ & 3.0 & 3.0 & \\
\end{longtable*}
\end{center}

\end{document}